\documentclass[hidelinks,letterpaper, 10 pt, conference]{ieeeconf}
\IEEEoverridecommandlockouts
\usepackage{cite}

\usepackage{algorithmic}
\usepackage{graphicx}
\usepackage{textcomp}
\usepackage{xcolor}
\def\BibTeX{{\rm B\kern-.05em{\sc i\kern-.025em b}\kern-.08em
    T\kern-.1667em\lower.7ex\hbox{E}\kern-.125emX}}
\usepackage{float}
\usepackage{subfig}	
\usepackage[font=small]{caption}
\usepackage{amssymb}
\usepackage{hyperref}
\usepackage{amsmath,amssymb,amsfonts}

\usepackage{amsthm}
\usepackage{color,soul}
\usepackage{caption}
\newtheorem*{definition*}{Definition}
\newtheorem{prop}{Proposition}
\newtheorem{theorem}{Theorem}

\def\baselinestretch{0.99}

\title{\bf A Control Theoretic Study on Omnidirectional MAVs with Minimum Number of Actuators and No Internal Forces at Any Orientation}
\author{Ahmed Ali$^{*,\dagger}$, Chiara Gabellieri$^{*}$, Antonio Franchi$^{*,\dagger}$
\thanks{$^*$ Robotics and Mechatronics Department, Electrical Engineering,  Mathematics, and Computer Science (EEMCS) Faculty, University of Twente, 7500 AE Enschede, The Netherlands. {\footnotesize ahmed.ali@utwente.nl, c.gabellieri@utwente.nl, a.franchi@utwente.nl}}
\thanks{$^\dagger$Department of Computer, Control and Management Engineering, Sapienza University of Rome, 00185 Rome, Italy. {\footnotesize ali.1987837@studenti.uniroma1.it, antonio.franchi@uniroma1.it}}\thanks{This work was partially funded by the Horizon Europe research agreement no. 101120732 (AUTOASSESS) and by the NWO OTP project AVIATOR.}}
\IEEEoverridecommandlockouts 
\begin{document}
\maketitle
\begin{abstract}
We propose a new multirotor aerial
vehicle class of designs composed of a multi-body structure in which a main body is connected by passive joints to links equipped with propellers. We have investigated some instances of such class, some of which are shown to achieve omnidirectionality while having a minimum number of inputs equal to the main body Degrees of Freedom DoF's, only uni-directional positive thrust propellers, and no internal forces generated at steady state. After dynamics are derived following the Euler-Lagrange approach, an I/O dynamic feedback linearization strategy is then used to show the controllability of any desired pose with stable zero dynamics. We finally verify the developed controller with closed-loop simulations.
\end{abstract}
\section{Introduction}
Over the past years, MAVs, short for Multirotor Aerial Vehicles, have received a great deal of research interest for their wide applicability in various industrial fields. According to the number of propellers, their configuration with respect to the main body of the vehicle and the vehicle’s geometrical shape, MAVs can be classified into different
categories such as Birotors, Trirotors, Quadrotors, Hexarotors, etc. An elaborate classification of this type of aerial vehicles is reported in \cite{Taxonomy}. In the traditional structure, all propellers are coplanar and collinear, meaning that they are mounted fixed on the same plane and all point in the same direction. This design has a serious limitation
obstructing the goal of making MAVs multi-purpose and cross-application as the available set of feasible motions is restricted by the structural design itself. This can be attributed to the fact that the translational dynamics of the vehicle are coupled with its rotational dynamics. As a consequence, only the position trajectories and
orientation around the z-axis trajectory can be arbitrarily assigned, thus omnidirectionality, as properly defined in \cite{Understanding}, is not present in those traditional MAV designs.

Various types of solutions have been suggested in literature in recent years to equip the vehicle with the omnidirectional property. A comprehensive review of fully-actuated MAVs is presented in \cite{review}. The common feature of these solutions seems to be that they decouple the translational and rotational parts of the dynamics by allowing an independent application of the total thrust force and the generated moment w.r.t. the main body's center of mass (CoM). This is accomplished by diversifying the tilting of the propellers w.r.t. the main vehicle's body, either through employing a  fixed-tilt mounting or a variable-tilting servomechanism.

Examples of fixed-tilt designs are abundant. The tilt angle is usually obtained as a solution to an optimization problem. In \cite{Theortical}, it is reported that a MAV with fixed-tilt unidirectional positive thrust propellers must have at least 7 of such propellers to be omnidirectional. A vehicle, named $O^7_+$, built based on the theory developed therein is described here \cite{O7+}. This option is also explored in \cite{EightRot}, where the ETH’s Omnicopter, a vehicle with 8 reversible propellers, is presented. Similarly, ODAR \cite{ODAR}, a MAV with 6 asymmetrically aligned and reversible propellers is designed. Moreover, the geometry by which the propellers are configured can be exploited. This is the main design principle behind the Lynchpin’s structure in \cite{Lynchpin}, which is also a vehicle with 6 reversible propellers.

Varying the propeller orientation with respect to the main body is the other widespread technique. The authors in \cite{Binbin} utilize a combination of Double-Gimbal Thrust Modules (DGTMs), each actuated by servo motors, for thrust vectoring. Voliro, a 12-input hexarotor with actuated tilting of the propellers, is introduced in \cite{Voliro}. Other MAVs with 12 inputs that incorporate servomechanism for tilting are detailed in DRAGON \cite{MultiLink_Dragon} and S3Q \cite{PassiveQuad}. An interesting structure composing a MAV capable of switching between two operating modes based on the type of the assigned trajectory, using only one additional servo motor, is developed in \cite{Morphing}. In \cite{HoloCopter}, a vehicle with 4 independently actively-tillable propellers with 4 servomotors is described.

Despite their evident success, a careful review would reveal that all of these designs can be characterized by at least one of the following: \emph{over-actuation} because of using an number of inputs that is larger than the DoF's of the main body, use of \emph{reversible-thrust} propellers, and having \emph{internal forces} unnecessarily produced at steady state when hovering at most of the orientations. Over-actuation comes with increased weight and cost, while reversible-thrust propellers and internal forces at hovering lead to significant energy waste (especially the latter).

In this paper, we introduce a new design that can overcome all these shortcomings. Although our analysis is given assuming 2D settings, similar to \cite{2D}, this study is primarily aimed at illustrating the new concept and demonstrating its potential, paving the way for the future extension to the corresponding 3D case. 

The main contribution is to prove that it is possible to conceive a MAV, with a number of control inputs equal to the DoF's of the main body, that is able to control its own pose (Theorem~\ref{thm:controllability}) and to achieve omnidirectionality (Prop. 1), while not producing any internal forces at equilibrium and using only unidirectional positive thrust propellers. 

To the best of our knowledge, this is the first time that a concept possessing all these control theoretic properties together is presented.
If translated into the real world, all these properties would have the potential to result in omnidirectional MAV's  that are cheaper, easier to build and maintain, and consume significantly less energy. 

The paper is organized as follows: First, the schematic design is elaborated and the dynamic model is obtained in Sec.~\ref{sec:dyn-model}. This is followed by a discussion on omnidirectionality in Sec.~\ref{sec:omni}. Afterwards, we devote Sec.~\ref{sec:cont}, to the introduction of a theorem about I/O feedback linearizability and the construction of the controller. We conclude with numerical simulations, whose results are presented in Sec.~\ref{sec:sim}, of the closed-loop vehicle model under the developed controller.

\section{Dynamic Model}
\label{sec:dyn-model}

\begin{figure}
\begin{minipage}[c]{0.49\columnwidth}
    \centering
   \subfloat[\centering {Type }1]{{\includegraphics[width=1.03\textwidth]{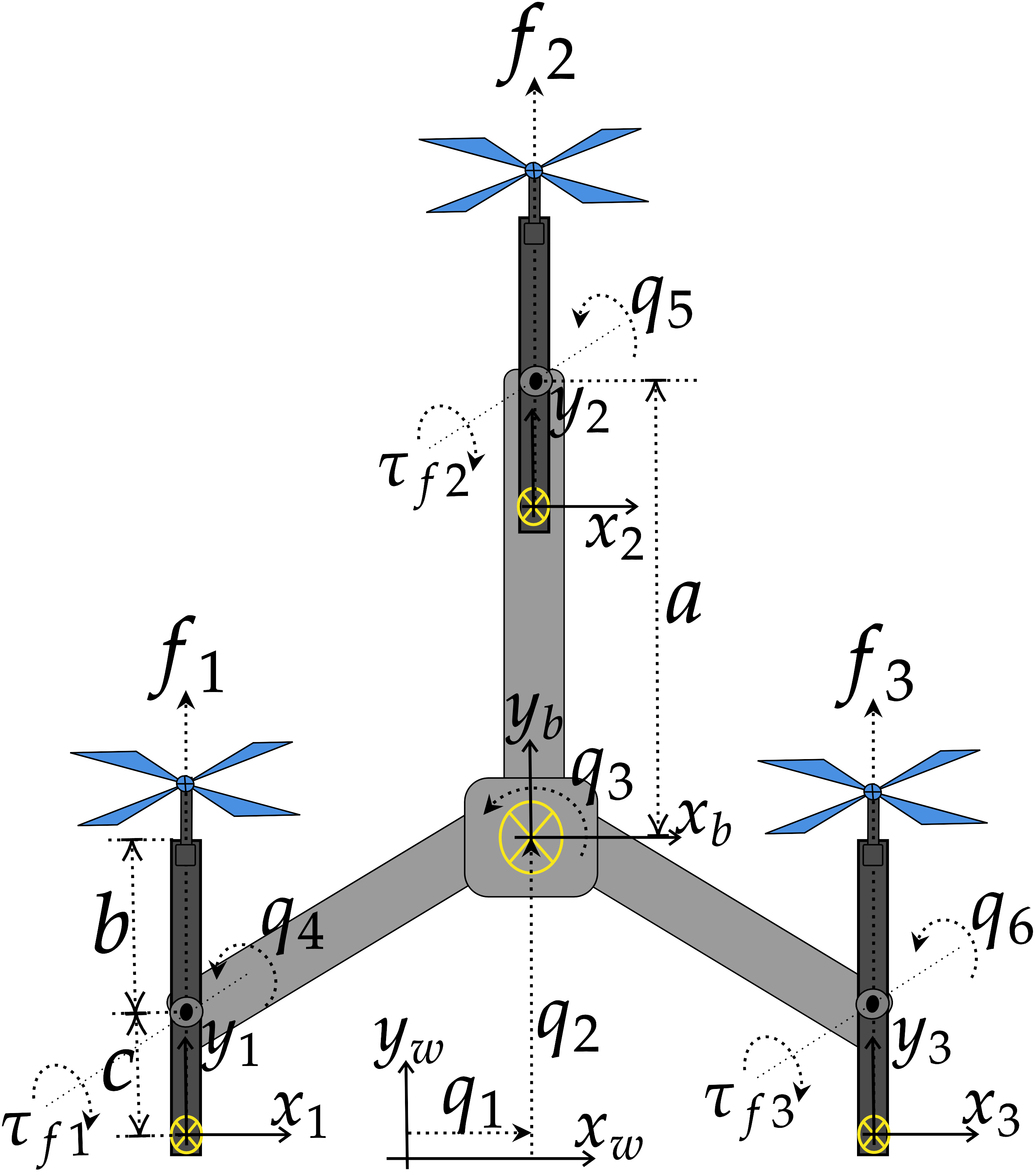} }}
\end{minipage}
\begin{minipage}[c]{0.49\columnwidth}   
       \subfloat[\centering {Type }2, Option 2: Via coupled-rotor propeller module]{\includegraphics[width=1\textwidth]{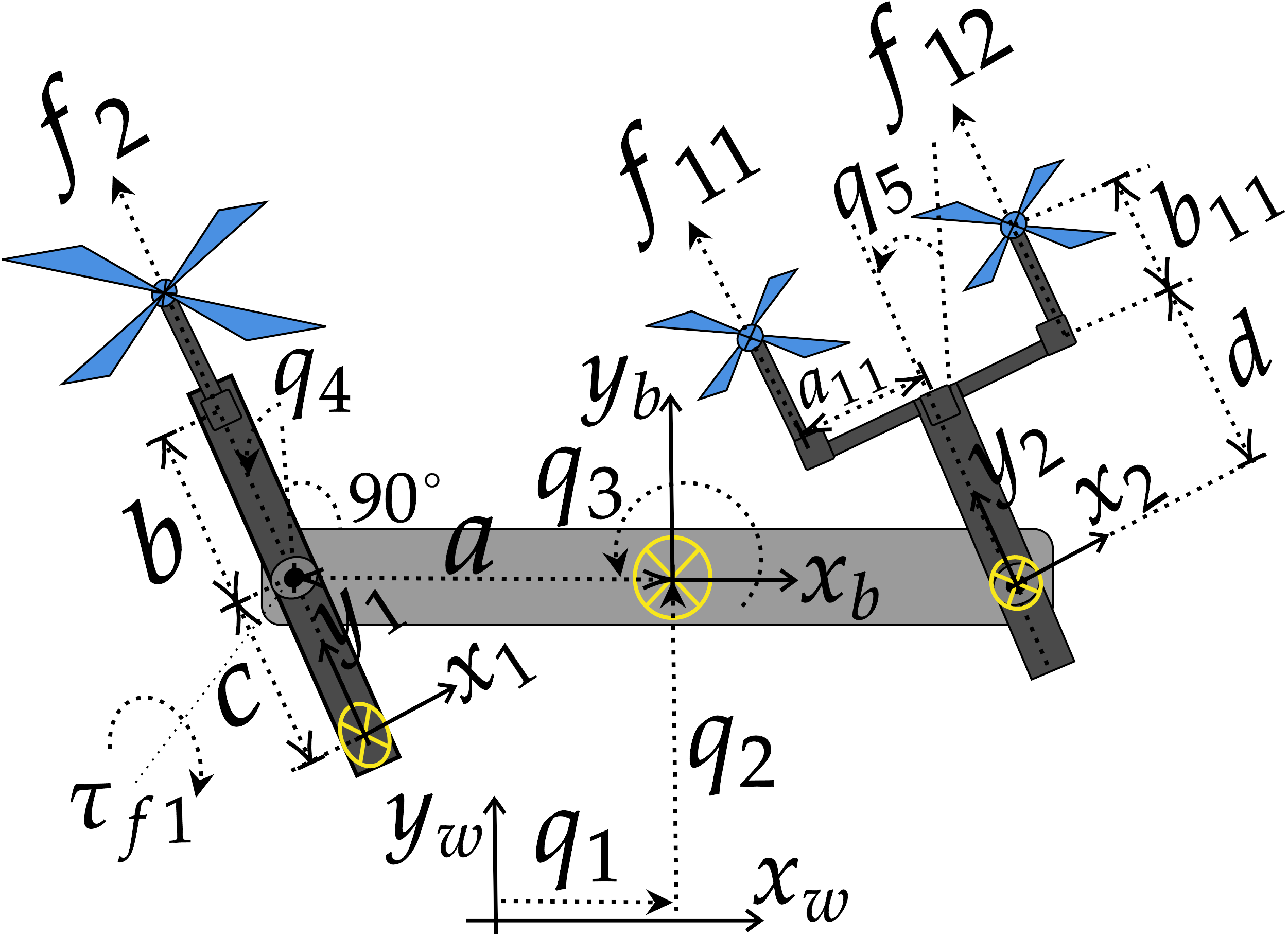}}\\
       \subfloat[\centering  {Type }2, Option 1: Via Servo motor]{{\includegraphics[width=1\textwidth]{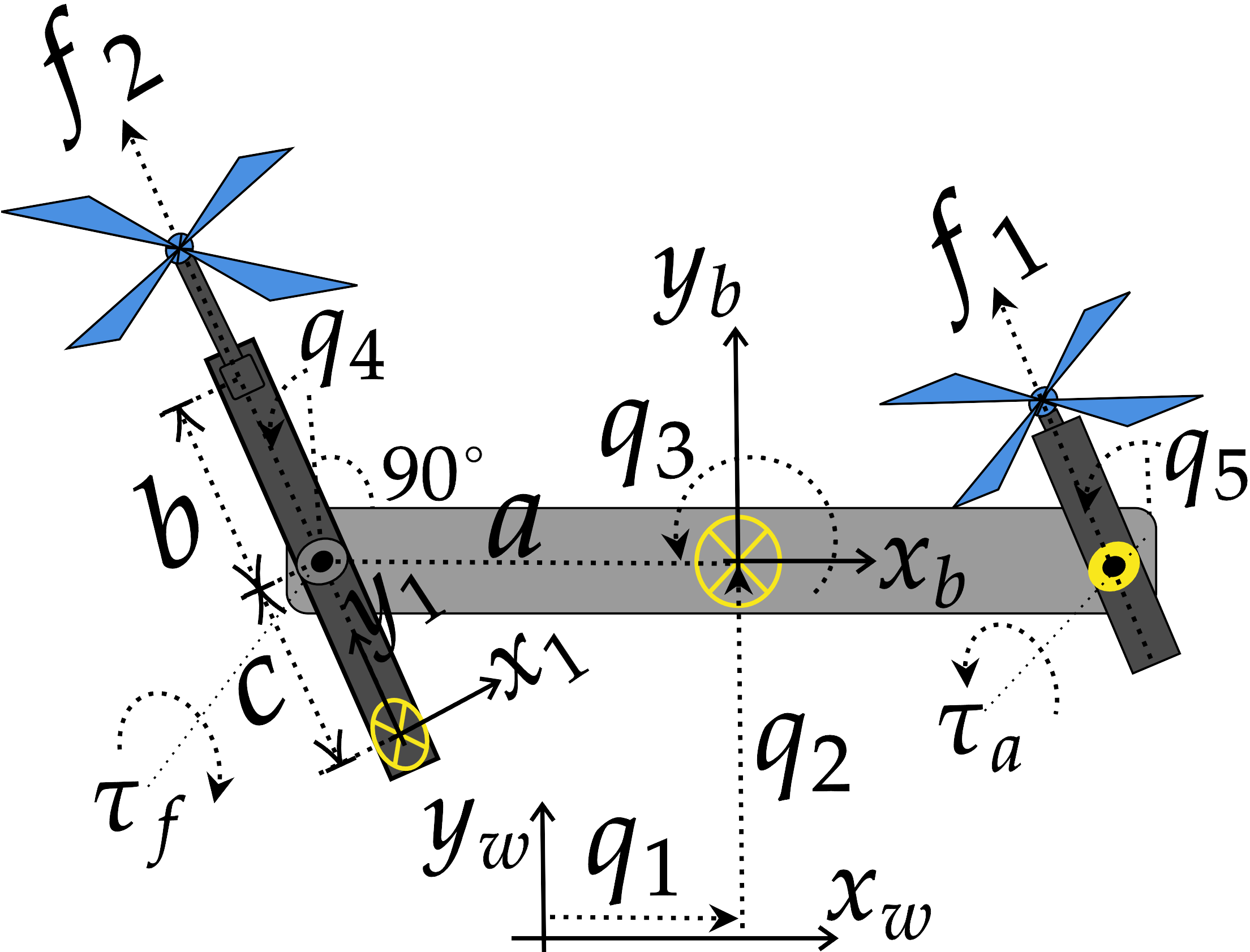} }}
\end{minipage}
    \caption{A schematic representation of instances in {Type }1 and 2, where $N$=3 and $N$=2, respectively. The non-inertial body frames are attached to each body's CoM. $f_i$ is the thrust of the propeller $i$. $\tau_a$ is the servo torque while $\tau_{f_i}$ denotes the friction at joint $i$.}
    \label{fig:designs1}
\end{figure}

Consider the vehicles in Fig.~\ref{fig:designs1}. They move on the 2D vertical plane. We differentiate between two designs throughout this text. They are sorted into these two {types} of vehicles:
\begin{itemize}
    \item \textbf{{Type }1}: The system consists of a main rigid body connected to $N$ rigid links, each of which carries a single propeller unit mounted on it, through $N$ passive, \textit{non-actuated} joints subject to viscous friction. The center of mass (CoM) of each link is located at a distance $c$ from the respective joint on the opposite side to the one in which the propeller is placed.
    \item \textbf{{Type }2}: This category spans designs distinct from {Type }1 in two ways: how one of the links is attached in the structure and the characteristics of that link. {Type }2 vehicle has at least one link connected to the main body through a \textit{moment-actuated} friction-less joint, either by the direct use of servo motor (Option 1), in which case this joint is active, or by utilizing a coupled-rotor propeller module (Option 2), which allows retaining the passive joint in the linkage. This link, denoted here by link $N$, must have its CoM located as close as possible to the corresponding joint to minimize the inertial couplings between the link and the remaining bodies. {The coupled-rotor propeller serves a similar purpose as the servo motor. It uses the differential thrust of the two propellers to enable the control of the moment around the joint rotation axis, hence the term `moment-actuated'}.
\end{itemize}
The vehicle has a configuration space $\mathbb{R}^2 \times (\mathcal{S}^1)^{N+1} \approx \mathbb{R}^{N+3}$ in which we define these generalized coordinates $\boldsymbol{q}$, 
\begin{align}
    \begin{split}
        \boldsymbol{q}&= \left(\begin{array}{cccccccc} x& y& 
        \phi &
        \theta_1 & \theta_2 & \theta_3&
        \cdots & \theta_{N+3}\end{array}\right)^T\\
        &=\left(\begin{array}{cccc} q_1& q_2& \cdots & q_{N+3}
        \end{array}\right)^T 
    \end{split}
\end{align}
 {where} $\left[x, y\right]^T \in
         \mathbb{R}^2$ and $\phi \in
         \mathcal{S}^1\approx \mathbb{R}$ are the position vector of the origin of the body-fixed frame of the main body and its orientation, respectively, expressed in the inertial world frame. The coordinate $\theta_i \in
         \mathcal{S}^1\approx \mathbb{R}$ denotes the $i$-th joint variable. To derive the dynamics, we follow the standard Euler-Lagrange (EL)  approach. After constructing the total Lagrangian function $L(\boldsymbol{q},\dot{\boldsymbol{q}})=T(\boldsymbol{q},\dot{\boldsymbol{q}})-U(\boldsymbol{q})$, where $T,\, U\in \mathbb{R}$ and $\dot{\boldsymbol{q}} \in T_{\boldsymbol{q}}\mathbb{R}^{N+3} $ are the total kinetic energy, the total potential energy and the generalized velocity vector of the system, respectively, we apply the EL equation,
\begin{equation} 
\frac{d}{d t} \frac{\partial L(\boldsymbol{q}, \dot{\boldsymbol{q}})}{\partial \dot{\boldsymbol{q}}}-\frac{\partial L(\boldsymbol{q}, \dot{\boldsymbol{q}})}{\partial \boldsymbol{q}}=\boldsymbol{Q}(\boldsymbol{q}, \dot{\boldsymbol{q}})
\end{equation}
{where} $\boldsymbol{Q} \in \mathbb{R}^{N+3}$ is the generalized force vector. The equations of motion can then be obtained, after some algebraic manipulations, in this form,
\begin{equation}
   \boldsymbol{M}(\boldsymbol{q}) \ddot{\boldsymbol{q}}+\boldsymbol{h}(\boldsymbol{q}, \dot{\boldsymbol{q}})+\boldsymbol{g}(\boldsymbol{q})=\boldsymbol{Q}(\boldsymbol{q}, \dot{\boldsymbol{q}}) 
   \label{EoM}
\end{equation}
{where} the generalized mass matrix $\boldsymbol{M}(\boldsymbol{q}) \in \mathbb{R}^{(N+3)\times{(N+3)}}$, the Coriolis and Centrifugal forces $\boldsymbol{h}(\boldsymbol{q}, \dot{\boldsymbol{q}}) \in \mathbb{R}^{N+3}$ and the gravitational force vector $\boldsymbol{g}(\boldsymbol{q}) \in \mathbb{R}^{N+3}$ are listed below,
\begin{align}
 \boldsymbol{M}=&\left(\begin{smallmatrix}
\boldsymbol{M}_{1}^{2 \times 2} & \boldsymbol{M}_{2}^{T} \\
\boldsymbol{M}_{2}^{(N+1) \times 2} & \boldsymbol{M}_{3}^{(N+1) \times(N+1)}
\end{smallmatrix}\right) \\
 \boldsymbol{M}_{2}=&\left(\begin{smallmatrix}
\sum\limits_{i=4}^{N+3} k_{1 i} \, c_{3 i l_i} & \sum\limits_{i=4}^{N+3}k_{1 i} \, s_{3 i l_i} \\
k_{1 4} \, c_{3 4 l_4} & k_{1 4}  \,s_{3 4 l_4} \\
\vdots & \vdots \\
k_{1(N+3)}  \,c_{3(N+3)l_{N+3}} & k_{1(N+3)} \, s_{3(N+3)l_{N+3}}
\end{smallmatrix}\right)\nonumber \\
\boldsymbol{M}_{3}=&\left(\begin{smallmatrix}
\sum\limits_{i=4}^{N+3} k_{2 i}+k_{3}+\sum\limits_{i=4}^{N+3} N k_{4 i}  \,s_{i}& \boldsymbol{M}_{4}^{T} \\
\boldsymbol{M}_{4}^{N \times 1}  & \boldsymbol{M}_{5}^{N \times N}
\end{smallmatrix}\right)\nonumber\\
\boldsymbol{M}_{4}=&\left(\begin{smallmatrix}
k_{24}+k_{44}  \,s_{4} \\
\vdots \\
k_{2(N+3)}+k_{4(N+3)}  \,s_{(N+3)}
\end{smallmatrix}\right) \nonumber
\end{align}
{where} $\boldsymbol{M}_{1}$ $=$ $\operatorname{diag}\left(m_{\text {tot}}\right)$ and $\boldsymbol{M}_{5}$$=\operatorname{diag}\left(k_{24},..., k_{2(N+3)}\right)$. We set $k_{1i}= d_i\,m_p,\, k_{2i}= I_p + d_i^2m_p,\,k_{3}= N m_p a^2+I_b,\,k_{4i}= a \, d_i\,m_p,\,m_{tot}=m_b+N\,m_p$, where $I_p$ and $m_p$ are the inertia moment around the z-axis of an attached link and its mass, respectively. Similarly, $I_b$ and $m_b$ represent the same parameters but for the main body. We use $s_i$, $c_i$, $s_{ijl_k}$ and $c_{ijl_k}$ as shorthand notations for $\sin(q_i)$, $\cos(q_i)$, $\sin(q_i+q_j\pm\theta_{l_k})$ and $\cos(q_i+q_j\pm\theta_{l_k})$, respectively, while $\theta_{l_k}$ is some constant angle dependent on the relative placing of the link-attached frame and the main body frame. $d_i$ and $a$ are the CoM location from the joint $q_i$ of the respective link, with $i =\{4,...,N+3\}$, and the length from the main body's CoM to the joint, respectively. For {Type }2, we always set $d_{N+3}=0$. Additionally, if Option 1 is selected, one can consider the dynamics of the last joint as $\ddot{q}_{N+3} = \tau_a$, where $\tau_a$ is the actuation torque, assuming negligible mass of its link $N$. 

The vectors $\boldsymbol{h}$ and $\boldsymbol{g}$ are given by, 
\begin{align}
\boldsymbol{h}=
\left(\begin{smallmatrix}
\sum\limits_{i=4}^{N+3} \pm k_{1i}\,s_{3il_i} {\left(\dot{q}_{3}+\dot{q}_{i}\right)}^2,
 \;
\sum\limits_{i=4}^{N+3} \pm k_{1i}\,c_{3il_i} {\left(\dot{q}_{3}+\dot{q}_{i}\right)}^2,
\end{smallmatrix}\right.\notag\\
\left.\begin{smallmatrix}
\sum\limits_{i=4}^{N+3} \pm k_{4i}\,c_{i}\dot{q}_{i}{\left(2\dot{q}_{3}+\dot{q}_{i}\right)},
\;\pm k_{44}\,c_{4}\dot{q}_{3}^2,
\;
\cdots,
\;
\pm k_{4(N+3))}\,c_{(N+3)}\dot{q}_{3}^2
\end{smallmatrix}\right)^T
\end{align}
\begin{align}
        \boldsymbol{g}&=\left(\begin{smallmatrix}
0,\;
g\,m_{tot} ,\;
\sum\limits_{i=4}^{N+3} \pm g\,k_{1i}\,s_{3il_i} ,\;
\pm g\,k_{14}\,s_{34l_4} ,\;
\end{smallmatrix}\right.\notag\\
&\left.\begin{smallmatrix}
\cdots,\;
\pm g\,k_{1(N+3)}\,s_{3(N+3)l_{(N+3)}}\end{smallmatrix}\right)^T 
\end{align}
{where} $g=9.81 m/s^2$. While those quantities are indeed the same for both {Types}, $\boldsymbol{Q}$ is different. It is given by,
\begin{equation}
    \boldsymbol{Q}=m\,\boldsymbol{e}_{N+3}\,\tau_a+ \sum^{N}_{j=1}\Biggl\{ -\boldsymbol{e}_p\,b_{f_p}\,\dot{q}_{p}+\sum^{k}_{i=1}{\boldsymbol{J}_{f_i}^T\,^{W}\boldsymbol{R}_{j}\boldsymbol{f}_i}\Biggr\}
\end{equation}
{where} $k$, $m$ and $b_{f_p}$ are the number of Cartesian forces acting on the link $j$, a binary indicator taking a value of either 0 for passive or 1 for active joint and the friction coefficient at joint $p$, respectively, with $p=j+3\text{ and } p\neq N+3$. Let $\boldsymbol{e}_l \in \mathbb{R}^{N+3}$ be a zero column vector with $1$ in the $l$-th place. $^{W}\boldsymbol{R}_{j} \in \mathcal{SO(\text{2})} \approx \mathbb{R}^{2\times2}$ represents the rotation matrix from the $j$-th link-attached frame to the world frame, whereas $\boldsymbol{J}_{f_i}$ is the jacobian of the position vector from the origin of the world frame to the point of application of force $\boldsymbol{f}_i$. Thus, for {Type }1, this yields the expression $\boldsymbol{Q}^1$, 
\begin{equation}
    \boldsymbol{Q}^1= \left(\begin{array}{c}
           f_{1}\,s_{34l_1}+f_{2}\,s_{35l_2}+\cdots+f_{N}\,s_{3(3+N)l_N}
           \\ f_{1}\,c_{34l_1}+f_{2}\,c_{35l_2}+\cdots+f_{N}\,c_{3(3+N)l_N}\\ \pm a\,f_{1}\,c_4\pm a\,f_{2}\,c_5\pm \cdots \pm a\,f_{N}\,c_{3+N}\\-\boldsymbol{B}_{f_{N\times N}} \bigl(\,\dot{q}_{4}\,\cdots\,\dot{q}_{N+3}\,\bigr)^T \end{array}\right)
           \label{Q1}
\end{equation}
While for {Type }2, we obtain $\boldsymbol{Q}^2$,
\begin{equation}
  \boldsymbol{Q}^2= 
        \left(\begin{array}{c} f_{1}\,s_{34l_1}+\cdots+u_{N_s}\,s_{3(3+N)l_N}\\ f_{1}\,c_{34l_1}+\cdots+u_{N_s}\,c_{3(3+N)l_N}\\ \pm a\,f_{1}\,c_4\pm\cdots\pm a\,u_{N_s}\,c_{3+N}\pm a_{11}\,(1-m)u_{N_d}\\-\boldsymbol{B}_{f_{(N-1)\times (N-1)}} \bigl(\,\dot{q}_{4}\,\cdots\,\dot{q}_{N+2}\,\bigr)^T \\ (1-m)\,a_{11}\,u_{N_d}+m\,\tau_a \end{array}\right)  
\end{equation}
{where} $u_{N_s}$ and $u_{N_d}$ are the sum and the difference of the coupled-rotor propeller thrusts, respectively, and $a_{11}$ is half the distance between its two rotors. $\boldsymbol{B}_{f}=\operatorname{diag}\left(b_{f_p}\right)$. The inputs can be taken as $\boldsymbol{u}^1 \in \mathbb{R}^{N}$ and $\boldsymbol{u}^2 \in \mathbb{R}^{N+1}$ with:
\begin{align}
&\boldsymbol{u}^1=\boldsymbol{f}_{N\times1}
   & \boldsymbol{u}^2=\left(\begin{array}{c}
         \boldsymbol{f}_{(N-1)\times1} \\
        (1-m)u_{N_s}+m\,f_N \\
        (1-m)u_{N_d}+m\,\tau_a 
    \end{array}\right) 
\end{align}
{where} $\boldsymbol{f}$ is a column vector of thrust magnitudes $f_i$. 

Equations~\eqref{EoM} can be put in the state space form of control-affine nonlinear systems. Let $i = \{1,2\}$ identify the {Type }selected and  $\boldsymbol{x}^i=\left( \begin{array}{cc}
    \boldsymbol{x}^i_1& \boldsymbol{x}^i_2
\end{array}\right)^T=\left( \begin{array}{cc}
   \boldsymbol{q}^i& \dot{\boldsymbol{q}}^i
\end{array}\right)^T \in \mathbb{R}^{2\, (N+3)}$ be the state vector, hence the state equations take the form,
\begin{equation}
    \begin{aligned}
 \dot{\boldsymbol{x}}^i&=
\boldsymbol{F}^i(\boldsymbol{x}^i)+\left(\begin{array}{c}
\boldsymbol{0}_{(N+3)\times(N+i-1)} \\
\boldsymbol{M}^{-1}\left(\boldsymbol{x}^i_1\right)\boldsymbol{J}_{Q^i}\left(\boldsymbol{x}_1^i\right)
\end{array}\right) \boldsymbol{u}^i
\\
& =\boldsymbol{F}^i(\boldsymbol{x}^i)+\boldsymbol{G}^i(\boldsymbol{x}_1^i) \boldsymbol{u}^i\\
\end{aligned}
\label{SS}
\end{equation}
 We let $\boldsymbol{J}_{Q^i} \in \mathbb{R}^{(N+3) \times (N+i-1)}$ denote the Jacobian of the generalized force $\boldsymbol{Q}^i$ with respect to the input $\boldsymbol{u}^i$. $\boldsymbol{F}^i$ and $\boldsymbol{G}^i$'s columns are smooth vector fields defined on $\mathbb{R}^{2\, (N+3)}$.

We conclude this section by computing the manifold embedded in the state space over which the vehicle reaches an equilibrium state. Let an equilibrium state be expressed by $\boldsymbol{x}^i_d=\left( \begin{array}{cc}
    \boldsymbol{x}^i_{1d}& \boldsymbol{0}
\end{array}\right)^T$, at which state we have, $\dot{\boldsymbol{x}}^i\equiv\boldsymbol{0}$. To find this subset of the state, the static balance equation is solved: $
\boldsymbol{g}(\boldsymbol{q}_d)-\boldsymbol{Q}^i(\boldsymbol{q}_d)=0
$. An admissible solution is found by setting the equilibrium inputs $\boldsymbol{u}^1_{d}$ and $\boldsymbol{u}^2_{d}$ as, 
\begin{align}
    &\boldsymbol{u}^1_{d}=\frac{g\,m_{tot}}{N} \,\boldsymbol{r}_{N\times1},&\boldsymbol{u}^2_{d}=\left(\begin{array}{c} \frac{g\,m_{tot}}{N} \,\boldsymbol{r}_{(N-1)\times1}\\0 \end{array}\right)
\end{align}
{where} $\boldsymbol{r}=(\,\,1\,\,\cdots\,\,1\,\,)^T$. For both {Types}, the equilibrium configuration attains this form,
\begin{equation}
    \boldsymbol{x}^i_{1d}= 
\Bigl(\, \, x_d\quad y_d\quad\phi_d\quad-\phi_d\mp\theta_{l_i}\,\, \,\,\cdots\,\,\, \,-\phi_d\mp\theta_{l_N} \, \,\Bigr)^T
\end{equation}
{where} the triplet $(x_d,y_d,\phi_d)$ corresponds to an arbitrary desired equilibrium pose of the main body. Thus, the admissible equilibrium configuration set $\mathcal{D}^i_q$ and its corresponding state set $\mathcal{D}^i_x$, $\forall i \in \{1,2\}$, can be defined by
\begin{align}
\mathcal{D}^i_x=\Bigl\{&\boldsymbol{x}^i(t)\in \,\mathbb{R}^{2(N+3)},\,\boldsymbol{u}^i(t)\in \mathbb{R}^{N+i-1},\,t,\Bar{t}\in \mathbb{R}_{\geq0} \nonumber \\
   &\Bigr\rvert\boldsymbol{x}^i(t)=\left( \begin{array}{c}
    \boldsymbol{x}^i_{1d}\\\boldsymbol{0}
\end{array}\right),
\boldsymbol{u}^i(t)=\boldsymbol{u}^i_{d},\,\forall t\geq\Bar{t}\Bigr\}\end{align}
\begin{equation}
    \mathcal{D}^i_q=\Bigl\{\boldsymbol{q}^i(t)\in \,\mathbb{R}^{N+3},\,t,\Bar{t}\in \mathbb{R}_{\geq0} 
   \,\Bigr\rvert\boldsymbol{q}^i(t)= 
    \boldsymbol{x}^i_{1d},\,\forall t\geq\Bar{t}\Bigr\} \nonumber
\end{equation}
In other words, in equilibrium configurations, the vehicle produces a total thrust lifting the vehicle weight while all the propellers are aligned vertically and point upward. In these equilibria, the propellers do not generate any horizontal forces, resulting in the absence of internal forces.
\section{Omnidirectionality analysis}
\label{sec:omni}
In this section, we begin by recalling the definition of the omnidirectional property of a MAV bound to fly on the 2D vertical plane. Afterward, we verify that {Type }2 meets the necessary and sufficient conditions for omnidirectionality, as illustrated in \cite{Understanding}, unlike {Type }1. First, let us denote the total applied wrench on the {Type }$i$ main body CoM, referred to the world frame, by $\boldsymbol{W}^i \in \mathbb{R}^3$. It is comprised from the total applied moment $W^i_{m} \in \mathbb{R}$ and force $\boldsymbol{W}^i_{f} \in \mathbb{R}^2$. This wrench is a subset of $\boldsymbol{Q}^i$. For the moment component, we have $W^i_{m}= Q^i_{3}$, while the force component is given by,
\begin{equation}
    \boldsymbol{W}^i=\left(\begin{array}{c}
      \boldsymbol{W}^i_{f} \\
     W^i_{m}
\end{array}\right),\, \boldsymbol{W}^i_{f}= \left(\begin{array}{c}
      {W}^i_{fx}\\ {W}^i_{fy}
\end{array}\right)=\left(\begin{array}{c}
      Q^i_{1}\\ Q^i_{2}
\end{array}\right)
\end{equation}
\begin{definition*}[Omnidirectional MAV \cite{Understanding}]
    Omnidirectionality is a term that describes a MAV which can change its total moment $W^i_{m}$ around any direction in $\mathbb{R}$, i.e. $rank\{\frac{\partial{W^i_{m}}}{\partial{\boldsymbol{u}^i}}\}=1$, and apply a non-zero total force $\overline{\boldsymbol{W}}^i_{f}\neq0$ in any direction in $\mathbb{R}^2$ while generating zero total moment $\overline{W}^i_{m}=0$. Moreover, if the vertical component of that force, called the lift force $\overline{W}^i_{fy}$, \textbf{always} counteracts the vehicle's weight, this MAV is fully omnidirectional (FOD). When this is true only in at least \textbf{one direction} in $\mathbb{R}^2$, it is said to be partially omnidirectional (POD). Otherwise, the MAV is not omnidirectional.
\end{definition*}
Based on this definition, these conditions are derived to check whether a MAV is omnidirectional: 
\begin{enumerate}
    \item The vehicle is fully-actuated w.r.t the task space, which is defined by $\mathbb{R}^3$ here in the 2D scenario. Hence it is required that $rank\{\frac{\partial{\boldsymbol{W}^{i}}}{\partial{\boldsymbol{u}^i}}\}=3$, and;
    \item The lift force of $\overline{\boldsymbol{W}}^i_{f}$ counteracts the weight at any or at least one direction in $\mathbb{R}^2$. Thus,  $\overline{W}^i_{fy}\geq g\,m_{tot}$. 
\end{enumerate}
\begin{prop}
At any equilibrium pose $\boldsymbol{q}^i_d \in \mathcal{D}^i_q$, {Type }1 vehicle is not omnidirectional while {Type }2 is fully omnidirectional.
\end{prop}
\begin{proof}
We lay out a constructive proof. By computing the {Type }1 wrench Jacobian of $\boldsymbol{W}^1$ w.r.t $\boldsymbol{u}^1$;$\frac{\partial{\boldsymbol{W}^{1}}}{\partial{\boldsymbol{u}^1}} \in \mathbb{R}^{3\times N}$,
\begin{equation}
 \frac{\partial{\boldsymbol{W}^{1}}}{\partial{\boldsymbol{u}^1}}=\left(\begin{array}{cccc} s_{34l_1} & s_{35l_2}&\cdots & s_{3(N+3)l_N}\\ c_{34l_1} & c_{35l_2}&\cdots & c_{3(N+3)l_N}\\ \pm a\,c_{4} & \pm a\,c_{5}&\cdots & \pm a\, c_{N+3} \end{array}\right)   
\end{equation} 
We find that it is rank-deficient at any $\boldsymbol{q}^1_d \in \mathcal{D}^1_q$ due to the vehicle not being able to change its acceleration in the lateral direction. Therefore, Condition 1 is violated, yielding {Type }1 vehicle not omnidirectional at equilibrium configurations.

For {Type }2, we have
$\frac{\partial{\boldsymbol{W}^{2}}}{\partial{\boldsymbol{u}^2}} \in \mathbb{R}^{3\times (N+1)}$,
\begin{equation}
    \frac{\partial{\boldsymbol{W}^{2}}}{\partial{\boldsymbol{u}^2}}=\left(\begin{array}{ccccc} s_{34l_1} &\cdots & s_{3(N+3)l_N}&0\\ c_{34l_1} & \cdots & c_{3(N+3)l_N}&0\\\pm a\,c_4 &\cdots &\pm a\,c_{3+N}&\pm (1-m)a_{11}\end{array}\right)
\end{equation}
Since we employ an actuated variable-tilting mechanism in {Type }2, \eqref{SS} are further derived twice w.r.t. time until the \textit{full allocation matrix} $\boldsymbol{F}_a$, relating variations in inputs to the applied wrench, is obtained \cite{Understanding}. For the instance where $N=2$, as depicted in Fig.~\ref{fig:designs1}, this matrix has this determinant,
\begin{equation}
    \operatorname{det}(\boldsymbol{F}_a)=-\frac{l_{3}\,c_4\,u_{N_s}}{l_{2}\,{c_4}^2+l_{1}}
\end{equation}
{where} $l_{i}$ $ \forall i$$ \in $$\{1,..,3\}$ is a positive constant. Hence, it can be deduced that {Type }2 is fully actuated for any configuration $\boldsymbol{q}^2_d \in \mathcal{D}^2_q-\{\boldsymbol{q}^2_d \in \mathcal{D}^2_q\, |\, \phi_d\neq pi/2 \}$, which, in this case, means that the vehicle is capable of changing the applied wrench in any direction in $\mathbb{R}^{3}$ at such $\boldsymbol{q}^2_d$ by varying the thrust acceleration and the moment at the joint $N+3$. 

Let us prove that Condition 2 is fulfilled for any pose of the main body. This implies that the vehicle can maintain the orientation of its main body constant while applying a force whose direction is arbitrary in $\mathbb{R}^{2}$ and lift amounts to the weight or higher. For Option 1, $\overline{W}^2_{m}=0$ requires $\pm u_{1}\,c_4\pm\cdots\pm \,u_{N_s}\,\overline{c}_{3+N}=0$, yielding $\pm \overline{c}_{3+N}=\frac{1}{u_{N_s}}(\mp u_{1}\,c_4\mp\cdots\mp \,u_{N-1}\,c_{2+N})$ where $u_{N_s}\neq0$. Note that the variable $q_{3+N}$ represents the orientation of propeller $N$, which is indeed directly actuated by servo motor torque $\tau_a$. This specific feature allows the nulling of the moment without restricting the set of admissible linear forces $\overline{\boldsymbol{W}}^2_{f}$ feasible at zero-moment. To see this, simply substitute this expression for $\overline{c}_{3+N}$ in $\overline{\boldsymbol{W}}^2_{f}\in \mathbb{R}^2$, yielding,
\begin{equation}
    \overline{\boldsymbol{W}}^2_{f}= \left(\begin{array}{c} u_{1} s_{34l_1}+\cdots+u_{N_s}f_s(u_{N_s, q_3,q_4,...,q_{2+N}})\\ u_{1} c_{34l_1}+\cdots+u_{N_s}\,f_c(u_{N_s, q_3,q_4,...,q_{2+N}})\\
    \end{array}\right) 
\end{equation}
{where} $f_s=sin(\cdot)$ and $f_c=cos(\cdot)$ with $(\cdot)=q_3+\theta_{l_N}+acos(\overline{c}_{3+N})$. This leads to $\overline{\boldsymbol{W}}^2_{f}$ having $N$ arbitrarily chosen thrust magnitudes in the set $\{\boldsymbol{u}^2\in \mathbb{R}^{N+1} | u_{N_s}\neq0\}$. Therefore, $\overline{\boldsymbol{W}}^2_{f}$ can assume any direction in $\mathbb{R}^{2}$ at any pose of the main body. Furthermore, as long as $N\geq2$, it is guaranteed that in any direction of $\overline{\boldsymbol{W}}^2_{f}$, the lift $\overline{W}^2_{fy}\geq g\,m_{tot}$, granting the property of FOD as the vehicle is capable of lifting its weight at any direction in $\mathbb{R}^{2}$ with zero moment. Similarly, for Option 2, when the moment vanishes, the set of feasible linear forces in $\mathbb{R}^{2}$ is obtained by setting $\pm \overline{u}_{N_d}=\frac{a}{a_{11}}(\mp u_{1}\,c_4\mp\cdots\mp \,u_{N_s}\,c_{3+N})$. This $u_{N_d}$ plays the same key role as $\tau_a$ in Option 1. Once again, $\overline{\boldsymbol{W}}^2_{f}$ can have both of its 2 components independently varied, while $N\geq2$, by the available $N-1$ thrusts and the control $u_{N_s}$, rendering possible an arbitrary change in both magnitude and direction of $\overline{\boldsymbol{W}}^2_{f}$. It follows that {Type }2 is FOD.\hspace*{\fill}
\end{proof}
\section{Controller Design}
\label{sec:cont}
We address the problem of nominally stabilizing a desired closed-loop pose of the main body of the vehicle $(x_d,y_d,\phi_d)$ by implementing a control strategy based on Input/Output Feedback Linearization. We present the main result in the following theorem whose proof contains the derivation of the control law. This result is applied to the case where the vehicle is only equipped with the minimum number of inputs w.r.t. the task space $\mathbb{R}^{3}$. This means that the vehicle has either 3 propellers or 2 propellers and 1 servo motor.

\begin{theorem}
\label{thm:controllability}
Let the output function be a sufficiently smooth mapping given by $\boldsymbol{h}^i(\boldsymbol{x}^i)= (x^i_{11} \, \,  x^i_{12} \,\,  x^i_{13})^T \in \mathbb{R}^3$ which is the main body pose. At any equilibrium $\boldsymbol{x}^{i}_d\in \mathcal{D}^i_x$, {Type }1 is neither statically nor dynamically I/O feedback linearizable whilst {Type }2, at least for the case where $N=2$, belongs to the {Type }of dynamically I/O feedback linearizable systems with stable zero dynamics.
\end{theorem}
\begin{proof}
The proof draws upon a well-known fundamental result in nonlinear control theory which states that the problem of I/O FBL, or alternatively, the non-interacting control problem is solvable at some state of the MIMO system by means of a diffeomorphism defining coordinate transformation and a \textit{static} state feedback \textit{iff} the system has a \textit{well-defined vector relative degree} at that state \cite{isidori2013nonlinear}. Moreover, if the system does not have such a vector relative degree, it might be possible to find a dynamic extension such that the combined system owns some uniform vector relative degree in an open neighbourhood of the point of interest. In that case, the system is dynamically I/O FB linearizable. 
    
For {Type }$i$ the relative degrees of individual outputs w.r.t $\boldsymbol{u}^i$ at $\boldsymbol{x}^{i}_d$ are $r_i=\{2,2,2\}$. If $r_i$ constitutes a \textit{vector} relative degree, the rank of the {Type }$i$ decoupling matrix $\boldsymbol{D}^i \in \mathbb{R}^{3\times (N+i-1)}$ evaluated at state $\boldsymbol{x}^{i}_d$, i.e. $\boldsymbol{D}^i(\boldsymbol{x}^{i}_d)$, is required to be \textit{constant} and equal $3$ in an open neighbourhood of $\boldsymbol{x}^{i}_d$. This matrix is derived from \eqref{SS} as follows,
    \begin{align}
        \boldsymbol{D}^i(\boldsymbol{x})=&\left(\begin{array}{ccc}
L_{\boldsymbol{g}_1} L_{\boldsymbol{F}}^{r_1-1} h_1(\boldsymbol{x}) & \cdots &L_{\boldsymbol{g}_{N+i-1}} L_{\boldsymbol{F}}^{r_1-1} h_1(\boldsymbol{x}) \\
L_{\boldsymbol{g}_1} L_{\boldsymbol{F}}^{r_2-1} h_2(\boldsymbol{x}) & \cdots & L_{\boldsymbol{g}_{N+i-1}} L_{\boldsymbol{F}}^{r_2-1} h_2(\boldsymbol{x}) \\
L_{\boldsymbol{g}_1} L_{\boldsymbol{F}}^{r_3-1} h_3(\boldsymbol{x}) & \cdots & L_{\boldsymbol{g}_{N+i-1}} L_{\boldsymbol{F}}^{r_3-1} h_3(\boldsymbol{x})
\end{array}\right)\nonumber\\
=& \, (\boldsymbol{M}^{-1}\left(\boldsymbol{x}^i_1\right)\boldsymbol{J}_{Q^i}\left(\boldsymbol{x}^i_1\right))^{'}
\end{align}
 {where} $L_{\boldsymbol{f}}^{i}\boldsymbol{h}$ is the $i$-th Lie derivative of $\boldsymbol{h}(\boldsymbol{x})$ w.r.t $\boldsymbol{f}(\boldsymbol{x})$. The superscript $'$ denotes the first 3 rows of the corresponding matrix. Since the matrix $\boldsymbol{M}^{-1}$ is always nonsingular, what remains is thus to investigate the rank of the Jacobian $\boldsymbol{J}_{Q^i}$ at $\boldsymbol{x}^{i}_d$. For {Type }1, it admits this form $\forall \boldsymbol{x}^{1}=\boldsymbol{x}^{1}_d\in \mathcal{D}^1_x$
\begin{equation}
\boldsymbol{J}_{Q^1}\left(\boldsymbol{x}^1_{1d}\right)=\left(\begin{array}{ccc}
0  & \cdots &0\\
1  & \cdots & 1\\
\pm\cos{(q_{d3})} & \cdots& \pm\cos{(q_{d3})} \\
&\boldsymbol{0}_{N\times N}&
\end{array}\right)
\end{equation}
Hence it is singular of rank 2 for any admissible equilibrium pose of the main body, resulting in a singularity in the matrix $\boldsymbol{D}^1(\boldsymbol{x}^{1}_d)$, which in turn deprives {Type }1 of the well-defined vector relative degree property at $\boldsymbol{x}^{1}_d$ and consequently exact static I/O decoupling. Furthermore, since elsewhere in an open neighbourhood of $\boldsymbol{x}^{1}_d$, $rank(\boldsymbol{J}_{Q^1})>2\,\,\, \,\forall N>2$, the rank of $\boldsymbol{D}^1$ is variable, rendering dynamic I/O FBL by a precompensator not possible \cite{Nijmeijer}. However, one may seek an \textit{approximate} I/O FBL by using a pseudoinverse of $\boldsymbol{D}^1(\boldsymbol{x}^{1}_d)$.

Let us now focus on {Type }2. The same argument shows that the decoupling matrix $\boldsymbol{D}^2(\boldsymbol{x}^2)$ of this {Type }is \begin{equation}
\left(\begin{array}{cc}
\boldsymbol{D}_{3\times N}& \mathbf{0}_{3\times 1}
\end{array}\right)
\end{equation}
which is, at least for the interesting square system when $N=2$, singular with a constant rank equal to 2 in an open neighbourhood of the equilibrium. As a result, it does not possess a vector relative degree at $\boldsymbol{x}^{2}_d$. However, since the rank is constant, the use of a dynamic extension algorithm (DEA) to obtain an extended system that possibly owns a vector relative degree is encouraged. For the case where $N=2$, running 2 iterations of DEA \cite{DEA} yields a combined system with 14 states $\boldsymbol{x}^{2}_E$, 4 of which are for this dynamic controller: 
$\dot{z}_{11}=z_{13}, \,
\dot{z}_{12}=z_{14}, \, 
\dot{z}_{13}=v_1, \, 
\dot{z}_{14}=v_2,
$ {where} $z_{11} \text{ and }z_{12}$ are the propeller thrusts $u_{1} \text{ and } u_{N_s}$, respectively, while $v_1,\, v_2 \text{ and } u_{N_d}\, (\text{or } \tau_a)$ are the new inputs of the combined system which are used to construct the I/O FB linearizing control law on that system. The new decoupling matrix of the extended system is full rank 3 as long as $u_{N_s}\neq0\text{ and }\phi_d\neq \pi/2$, i.e. $\boldsymbol{x}^{2}_E(t) \in \mathcal{L}_E = \{\boldsymbol{x}^{2}_E(t)\in \mathbb{R}^{14} | \,x^2_{14}(t)\neq \pm(2k+1)\pi/2,\, z_{12}(t)\neq0,\, \forall t\geq0,\, \forall k=0,1,2,...\}$, thus achieving a vector relative degree at $\boldsymbol{x}^{2}_d$ whose elements sum up to 12. Hence, the internal dynamics are two-dimensional. 

\begin{figure}[h!]
    \centering
   \subfloat[\centering Sim 1: Pose states]{{\includegraphics[width=0.48\columnwidth]{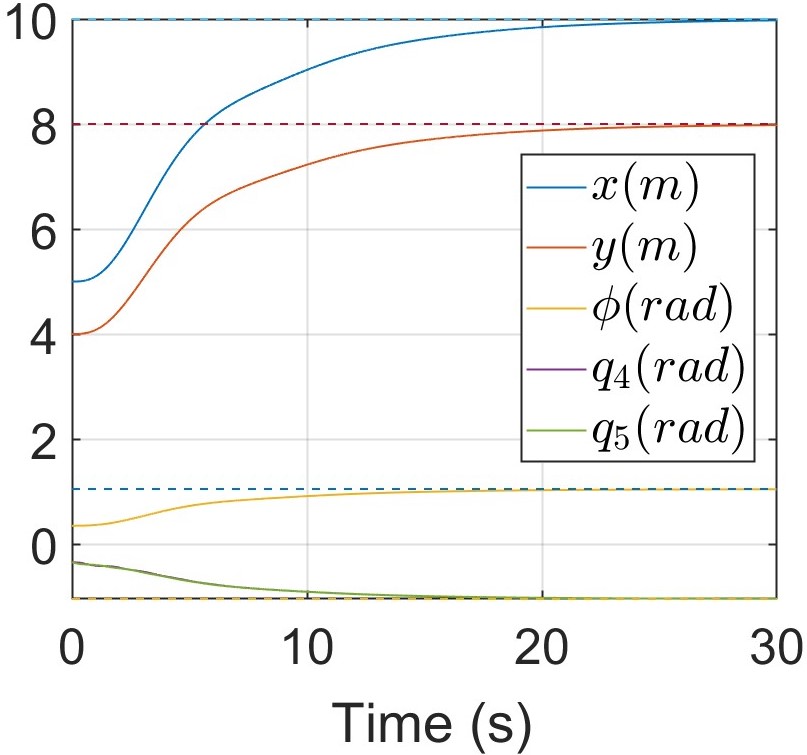} }}
       \subfloat[\centering Sim 2: Pose states]{{\includegraphics[width=0.48\columnwidth]{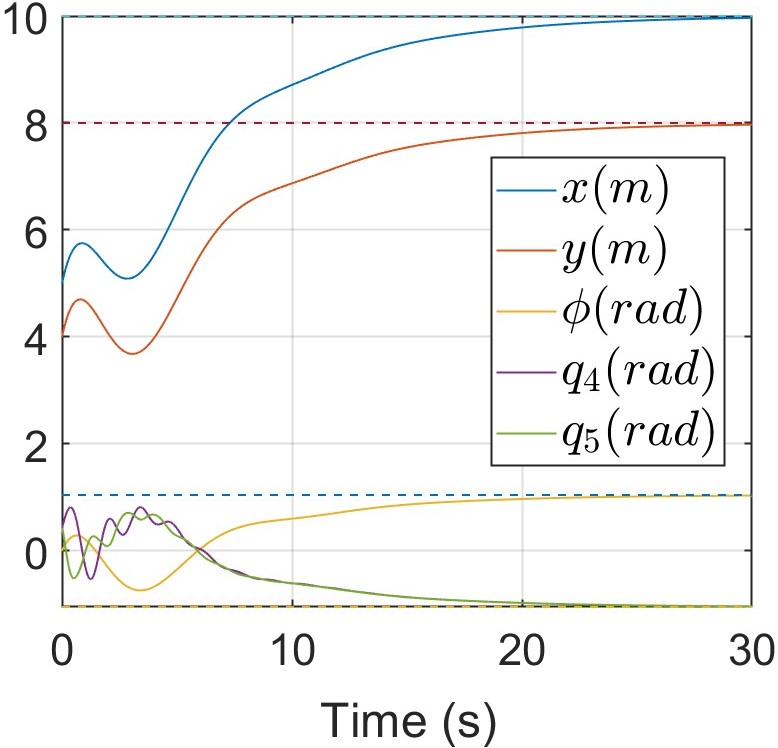} }}\\
    \subfloat[\centering Sim 1: Velocity states]{\includegraphics[width=0.48\columnwidth]{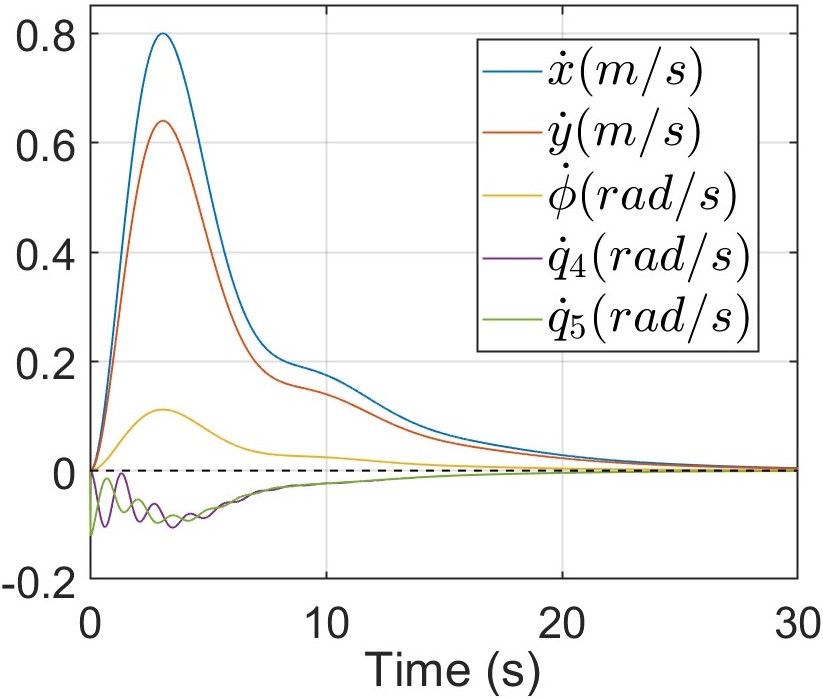}}
    \subfloat[\centering Sim 2: Velocity states]{\includegraphics[width=0.48\columnwidth]{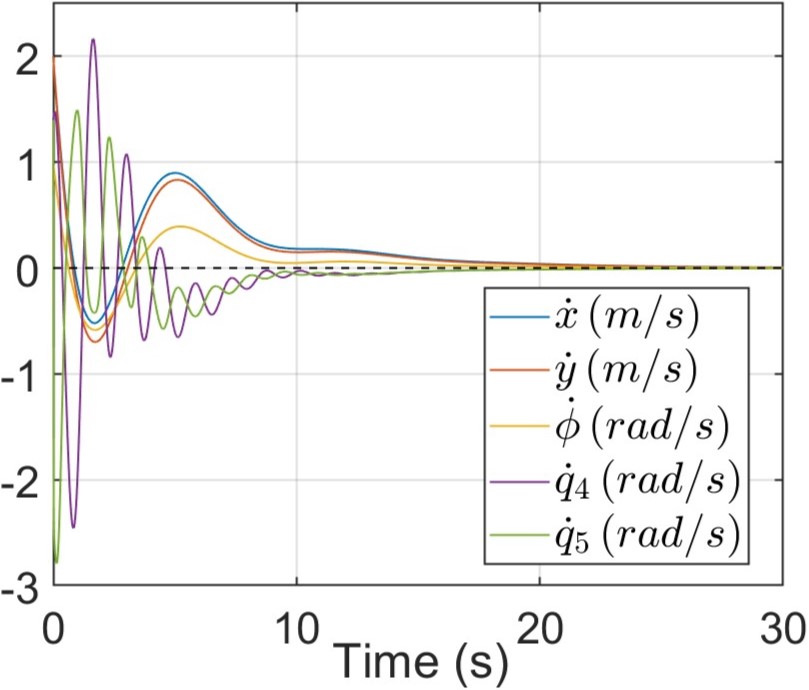}}\\
    \subfloat[\centering Sim 1: Control input]{{\includegraphics[width=0.48\columnwidth]{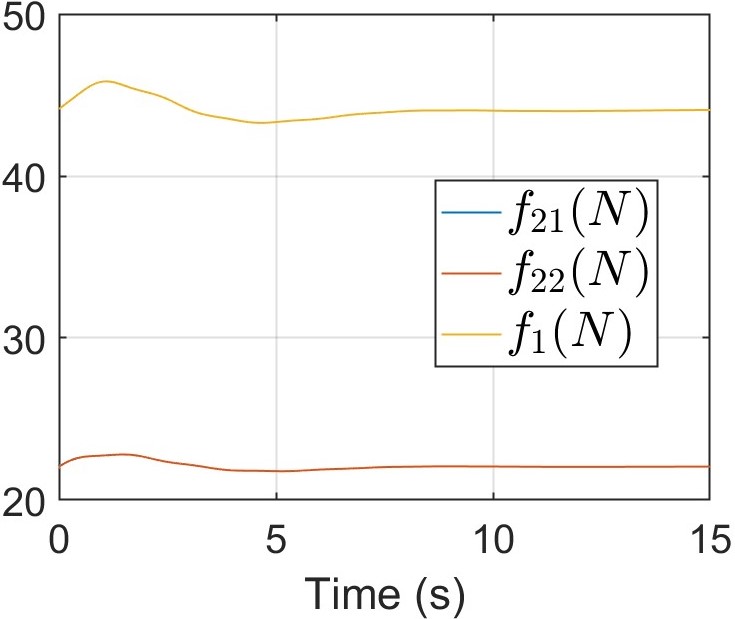} }}
    \subfloat[\centering Sim 2: Control input]{{\includegraphics[width=0.48\columnwidth]{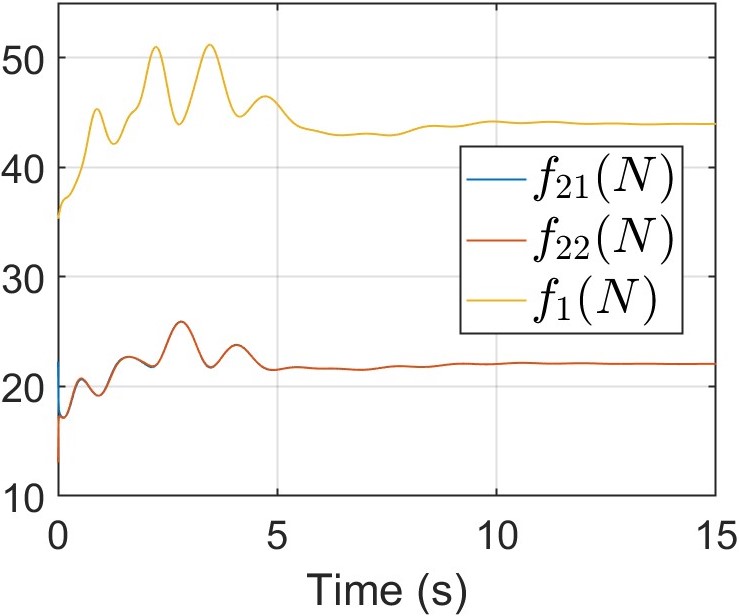} }}\\
   \subfloat[\centering ZD: Single initial state]{{\includegraphics[width=0.51\columnwidth]{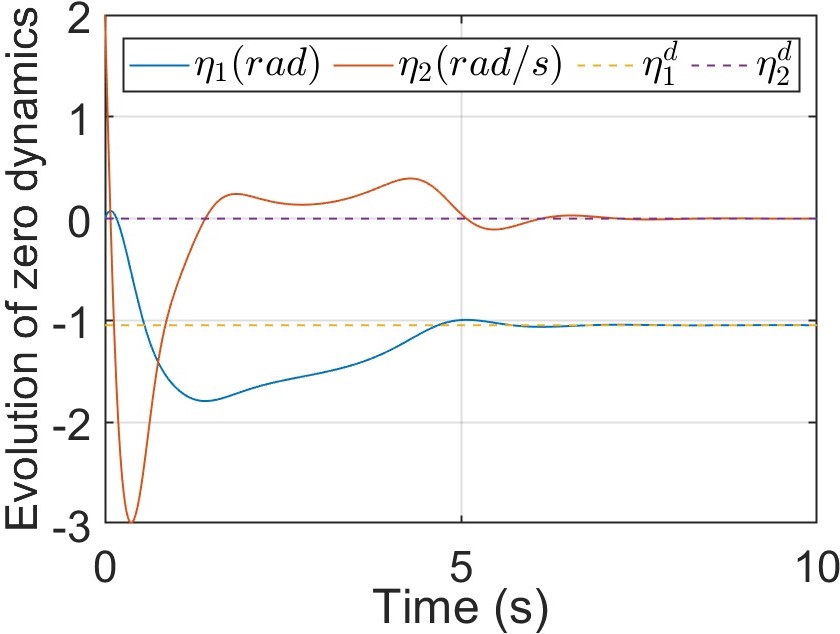} }}
    \subfloat[\centering ZD: Multiple initial states]{\includegraphics[width=0.48\columnwidth]{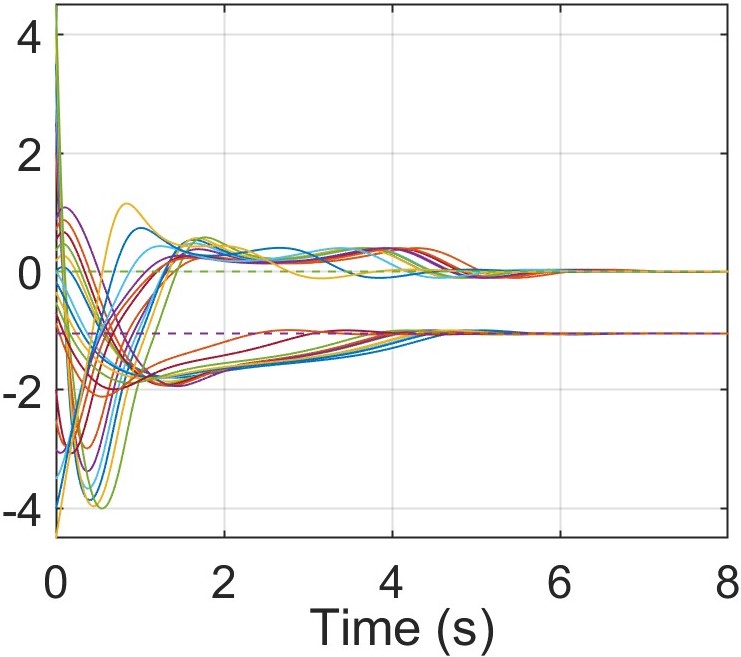}}
    \setlength{\belowcaptionskip}{-11pt}
    \caption{Evolution of states and control input is depicted. The initial state in {Sim 1} (a,c,e) is another admissible equilibrium $\in \mathcal{D}^2_x$ while in {Sim 2} (b,d,f) it is chosen randomly $\in \mathcal{L}_E$ in the vicinity of the desired equilibrium. Last row: zero dynamics evolution starts from a single initial condition (g) and multiple conditions (h).}
    \label{fig:SimAndZD}
\end{figure}

Moreover, the zero dynamics (ZD) can be derived in the transformed coordinates \cite{isidori2013nonlinear}. We find that this map $\boldsymbol{\Phi}(\boldsymbol{x}^{2}_E)=\bigl(\,\eta_1\quad \eta_2\,\bigr)^T=\bigl(\,{x}^2_{14}\quad x^2_{24}\,\bigr)^T:\,\mathcal{L}_E \longrightarrow \mathbb{R}^{2}$ is an admissible solution to the PDE: $L_{\boldsymbol{G}}\boldsymbol{\Phi}(\boldsymbol{x}^{2}_E)=0,\, \forall \boldsymbol{x}^{2}_E \in \mathcal{L}_E$, while making the diffeomorphism $\bigl(\,\boldsymbol{H}\quad \boldsymbol{\Phi}\,\bigr)^T:\,\mathcal{L}_E \longrightarrow \mathbb{R}^{14}$ full rank at $\forall \boldsymbol{x}^{2}_E \in \mathcal{L}_E$, with $\boldsymbol{H}=\bigl(\,\boldsymbol{h}\quad L_{\boldsymbol{F}}\boldsymbol{h} \quad L_{\boldsymbol{F}}^{2}\boldsymbol{h} \quad L_{\boldsymbol{F}}^{3}\boldsymbol{h}\,\bigr)^T$. $\boldsymbol{F}$ and $\boldsymbol{G}$ are the drift and input vector fields of the extended system, respectively. Interestingly, the zero dynamics states are those of the passive joint variables and their angular velocities, except the joint $N+3$. If the friction applied at those joints is removed, and the CoM of the corresponding links coincides with the joints, these equations for the ZD result,
\begin{align}
     \dot{\eta}_1=&{\eta}_2 \nonumber
     \\   \dot{\eta}_2=&l\,(c_4-\cos\left(\phi_d\right))
     \label{ZD1}
\end{align}
Otherwise, the ZD can generally be obtained in this form,
\begin{align}
     \dot{\eta}_1=&{\eta}_2 \label{ZD2}
     \\ 
     \dot{\eta}_2=
     &\tfrac{\left(l_{5}\,{\eta_2}^2-l_{7}\right)\,s_4\,c_4-\left(l_{10}+l_{9}\,\eta_2\right)\,s_4+l_{2}\,{c_4}^2+\left(l_{1}\,{\eta_2}^2-l_{3}\right)\,c_4+l_{6}-l_{8}\,\eta_2}{l_{5}\,{c_4}^2+l_{4}} \nonumber
\end{align}
{where} $l_{i}$ $ \forall i$$ \in $$\{1,..,10\}$ is a constant function of the vehicle parameters and $\phi_d$. $s_4$ and $c_4$ are $\sin(\eta_1)$ and $\cos(\eta_1)$, respectively. While it can be verified, relatively easily, that \eqref{ZD1} are unstable for any $\phi_d$, the case of \eqref{ZD2} is more analytically involved. For this reason, we resort to numerical simulation in Fig. \ref{fig:SimAndZD} to show that solutions of \eqref{ZD2} converge to the desired equilibrium $\bigl(\,\eta_1^d\quad \eta_2^d\,\bigr)^T=\bigl(\,-\phi_d\quad 0\,\bigr)^T,$ starting from some open neighbourhood around it.\hspace*{\fill}
\end{proof}

\section{Simulation Results}
\label{sec:sim}
Simulations were carried out in MATLAB/Simulink. In Fig.~\ref{fig:SimAndZD}, we present the results of some simulations of a {Type }2 vehicle under I/O FBL. The objective is regulating the platform pose to the desired value $(x_d,y_d,\phi_d)=(10\,\rm{m},8\,\rm{m}, 60^\circ)$. We show only Option 2 as Option 1 has an almost identical response. The numerical parameters are $m_{tot}=10\, \rm{kg},\, a=b=c=0.5\, \rm{m},\, a_{11}=0.1\,\rm{m},\, I_{b}=0.0095\,\rm{kg\cdot m}^2,\, I_{p}=0.002\,\rm{kg\cdot m}^2$. Different initial states of the vehicles are simulated.  Simulations of various other scenarios are available through videos at  
{\texttt{\small https://tinyurl.com/yc4t3fmr}}
\subsection{Robustness Analysis of Trajectory Tracking Scenario}
We expand the results on MAV belonging to Type 2 with further simulations. We consider here the problem of tracking a desired trajectory for the pose of the main platform when two types of non-idealities are introduced in the closed-loop system. In Sec.~\ref{sec:para}, we aim at investigating the effect of the parametric uncertainty, while the scenario where the vehicle is subject to external disturbance forces is demonstrated in Sec.~\ref{sec:dist}.

Suppose that the vehicle is tasked with tracking a position trajectory, which is described by a circle with a radius ${r_1=1\, \rm{m}}$ and center at $(\bar{x},\bar{y})=(5\, ,5)$ in the Euclidean space, while keeping its orientation constant at all times. Hence, we have these reference trajectories in the configuration space,
\begin{equation}
\left(\begin{array}{c}
      {x_d}(t)\\ {y_d}(t) \\ {\phi_d}(t) \\
\end{array}\right)=\left(\begin{array}{c}
     r_1 \cos(\theta(t)) + \bar{x}\\ r_1 \sin(\theta(t)) + \bar{y} \\  0 \\
\end{array}\right)
\end{equation}
where $\theta(t)=0.5 \, t$ determines the timing law of the trajectory. Furthermore, to better show the decoupling between the translational and rotational components of motion, the case where the desired orientation is time-varying in a sinusoidal manner, i.e. $\phi_d(t)=r_2 \sin(r_3\,t)$ with $r_2=80\rm{deg}$ and $r_3=30\,\rm{deg/s}$, is simulated for the nominal system. The simulation results in the nominal scenario are available here {\small\url{ https://tinyurl.com/yc4t3fmr}}.
\subsubsection{Robustness against parametric uncertainty}
\label{sec:para}
We have in total 7 parameters that characterize the vehicle model. 

\begin{figure}[h!]
        \resizebox{\columnwidth}{!}{\centering
   \subfloat[\centering Position tracking err. $e_{pos}(t)$ for $+\Delta$]{{\includegraphics[width=0.59\columnwidth]{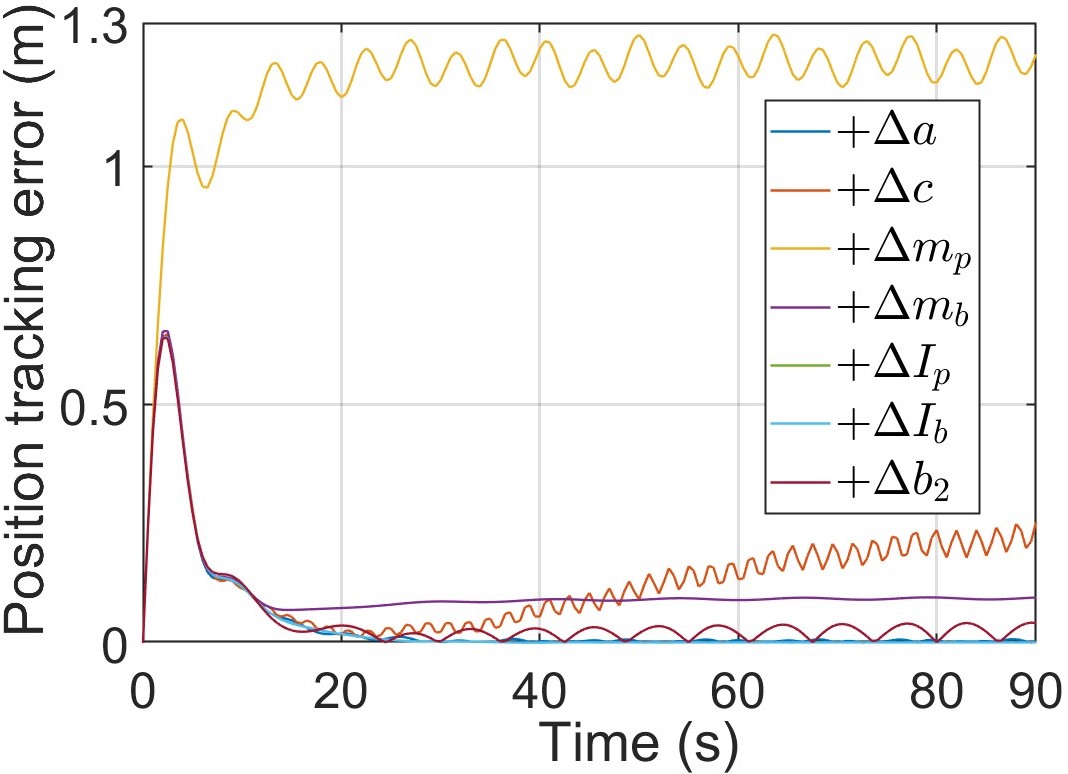}}}
       \subfloat[\centering Position tracking err. $e_{pos}(t)$ for $-\Delta$]{{\includegraphics[width=0.55\columnwidth]{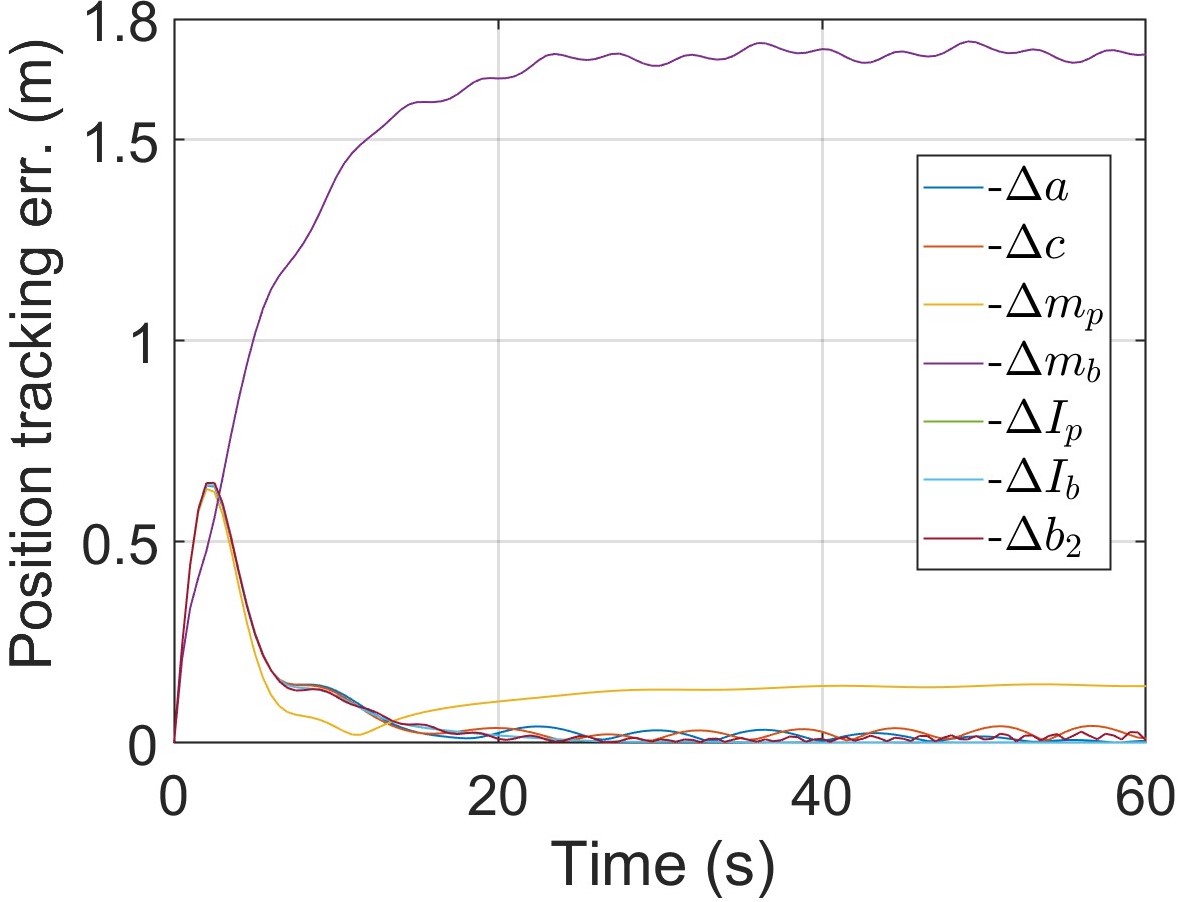}}}}\\
       \resizebox{\columnwidth}{!}{
    \subfloat[\centering Orientation tracking err. $e_{\phi}(t)$ for $+\Delta$]{{\includegraphics[width=0.55\columnwidth]{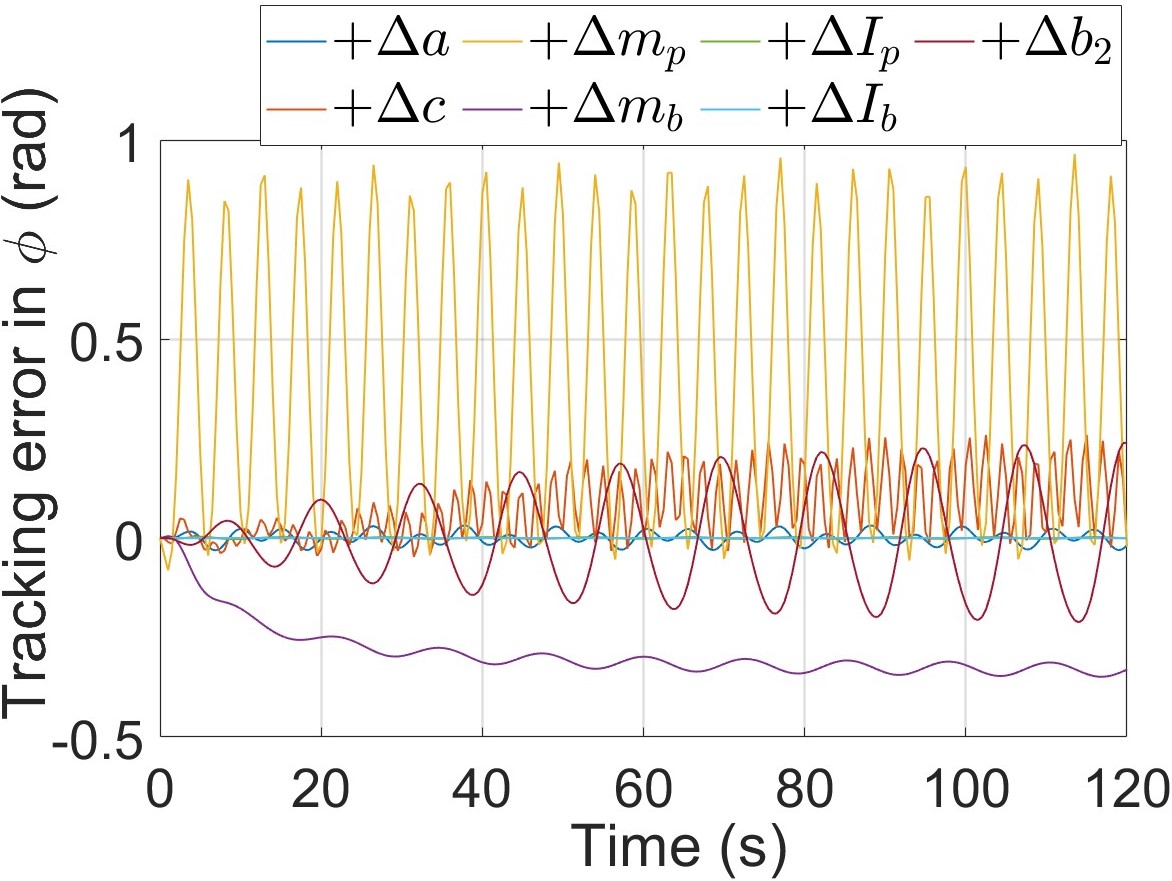}}}
    \subfloat[Orientation tracking err. $e_{\phi}(t)$ for $-\Delta$]{{\includegraphics[width=0.59\columnwidth]{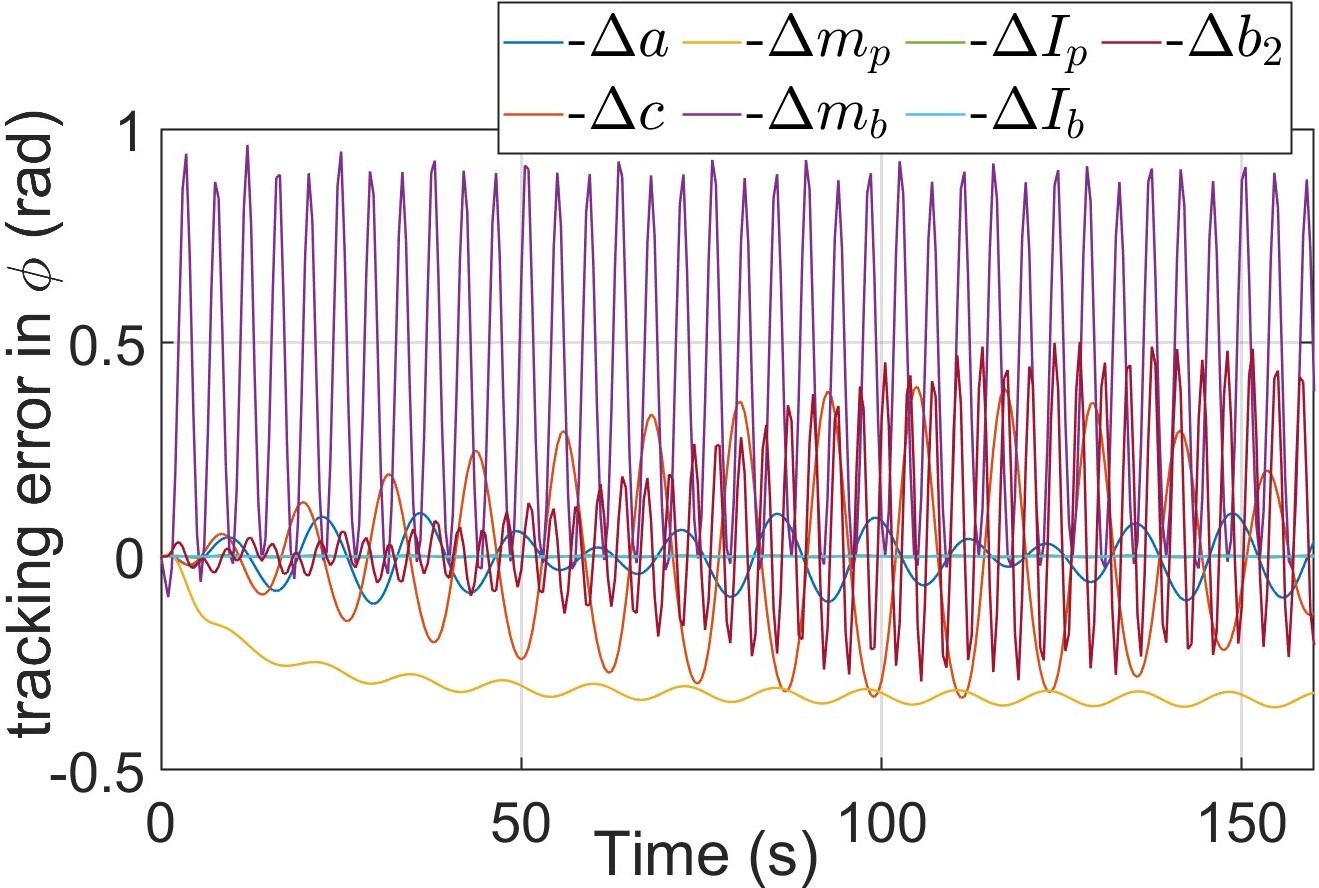}}} }
    \caption{Tracking errors of the orientation and position are shown for each maximum perturbation $\Delta p$ of each parameter in positive (a,c) and negative (b,d) magnitudes.}
    \label{Individ_pert}
\end{figure}

First, in order to determine the allowed variation range for each parameter, we follow a procedure in which all parameters in the plant are kept at their nominal values except one parameter which is the one to be perturbed. The value of this parameter $p$ is set according to
\begin{equation}
    \delta p= (1 \pm \Delta p) p
\end{equation}
where $\delta p$ and $p$ are, respectively, the perturbed value, used in the plant model, and the nominal value which the controller model utilizes. $\Delta p$ denotes the perturbation size. In Table~\ref{Tab:pert_sep}, nominal values of each parameter along with the allowed range of perturbations within which the closed-loop stability is preserved, and the vehicle does not encounter a singular configuration. In order to see how the tracking performance, taking the output tracking error as a metric, changes with perturbations of this kind, we plotted this error for both extremes of the range in Fig.~\ref{Individ_pert}. 

\begin{figure}[h!]
    \centering
   \centering {{\includegraphics[width=0.6\columnwidth]{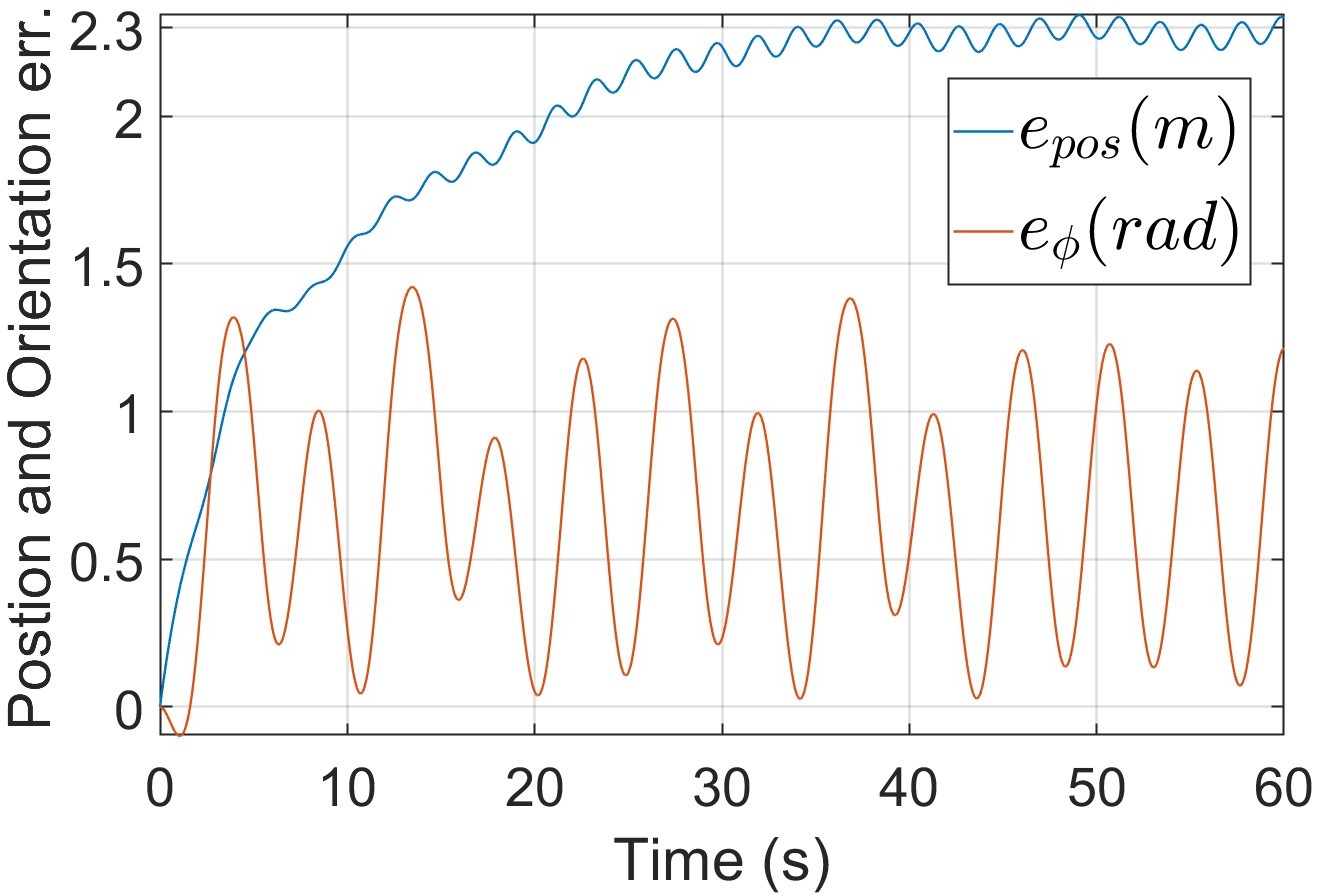} }}
    \caption{Evolution of the worst-case output error, within the uniformly sampled 1000 combined perturbations, is depicted.}
    \label{Worst_pert}
\end{figure}

Second, since we have a nonlinear dependence between the parameters and the closed-loop vector field, combined perturbations may have a totally different impact on the performance than separate perturbations. For this reason, we take 1000 samples uniformly distributed over the allowed ranges for each parameter,  and search for the combined perturbation, which produces the largest tracking error signal, in the sense of $L_\infty$ norm, of  the position and orientation coordinates. This is the worst-case perturbation in the sampled region for the respective error component. Let $\mathcal{P}$ represent the sampled region, then for every combined perturbation $\boldsymbol{\Delta P} \in \mathcal{P}$, this norm of the error is computed by, 

\begin{align}
      E_{pos}(\boldsymbol{\Delta P})=&\left\lVert e_{pos}(t)\right\rVert_\infty\\
E_\phi(\boldsymbol{\Delta P})=&\left\lVert \phi(t)\right\rVert_\infty, \text{ with $\phi_d(t)\equiv0$.}
\end{align}

where $e_{pos}(t)$ the tracking error of the position trajectory is evaluated at $\forall i \in T\equiv \{\text{simulation time instants\}}$ by $e_{pos}(i)= ||\boldsymbol{o}_a(i)-\boldsymbol{o}_d(i)||$ with $\boldsymbol{o}_a$ and $\boldsymbol{o}_d$ representing points on the actual and desired trajectories at instant $i$, respectively. The worst-case error norm, denoted by $\Bar{E}$, for these two components across this sampled perturbation region $\mathcal{P}$ can be thus given by,
\begin{equation}
    \bar{E}_i = \max_{\boldsymbol{\Delta P} \in \mathcal{P}} \{ E_i(\boldsymbol{\Delta P}) \}, \quad \forall i \in \{pos,\,\phi\}.
\end{equation}
The corresponding time evolution of worst-case errors are shown in Fig.~\ref{Worst_pert}. The perturbation sizes at which these errors occur are listed in Table~\ref{Tab:pert_comb}.
\begin{table}[h!]
\centering
\resizebox{\columnwidth}{!}{\begin{tabular}{|c|c|c|c|}
\hline
Parameter $p$ & Nominal & Range $\Delta p$ & \# Occurrences\\
\hline \hline
$a$ & 0.5 $(m)$ & [-20.8\%\,,\,+21.1\%] & 943\\
\hline
$c$ & 0.5 $(m)$ & [-10.28\%\,,\,+141\%] & 1384\\
\hline
$m_p$ & 2 $(kg)$ & [-0.149\%\,,\,+2.74\%] & 2347\\
\hline
$m_b$ & 5 $(kg)$ & [-3.15\%\,,\,+0.149\%] & 2151 \\
\hline
$b_2$ & 0.9 $(N.m.s/rad)$ & [-35\%\,,\,+7.15\%] & 769 \\
\hline
$I_p$ & $1.86\times10^{-3}$ $(kg.m^2)$ & [-900\%\,,\,+900\%] & 727 \\
\hline
$I_b$ & $9.5 \times 10^{-3}$ $(kg.m^2)$  & [-900\%\,,\,+900\%] & 547 \\
\hline
\end{tabular}}
\caption{Nominal value and allowed perturbation range for each parameter when they are perturbed separately.}
\label{Tab:pert_sep}
\end{table}
\begin{table}[h!]
\centering
\begin{tabular}{|c|c|c|}
\hline
Perturbation $\boldsymbol{\Delta P}$ & for position err. & for $\phi$-error \\
\hline \hline
$\Delta a$ & -16.90\% & 4.22\%  \\
\hline
$\Delta c$ &122.82\% & 1.31\% \\
\hline
$\Delta m_p$ & 1.93\% & 1.03\% \\
\hline
$\Delta m_b$ & -2.64\% & -2.34\% \\
\hline
$\Delta b_2$ & 7.06\% & -8.61\% \\
\hline
$\Delta I_p$ & 164.36\% & 80.52\% \\
\hline
$\Delta I_b$ & -704.07\%  & 42.79\% \\
\hline
\end{tabular}
\caption{Worst-case combined perturbations, within the sampled region, for each component of the output error.}
\label{Tab:pert_comb}
\end{table}
\subsubsection{Robustness against external disturbance}
\label{sec:dist}
\begin{figure}[t]
    \centering
   \subfloat[\centering Evolution of output error when fast-varying disturbances are included.]{{\includegraphics[width=0.8\columnwidth]{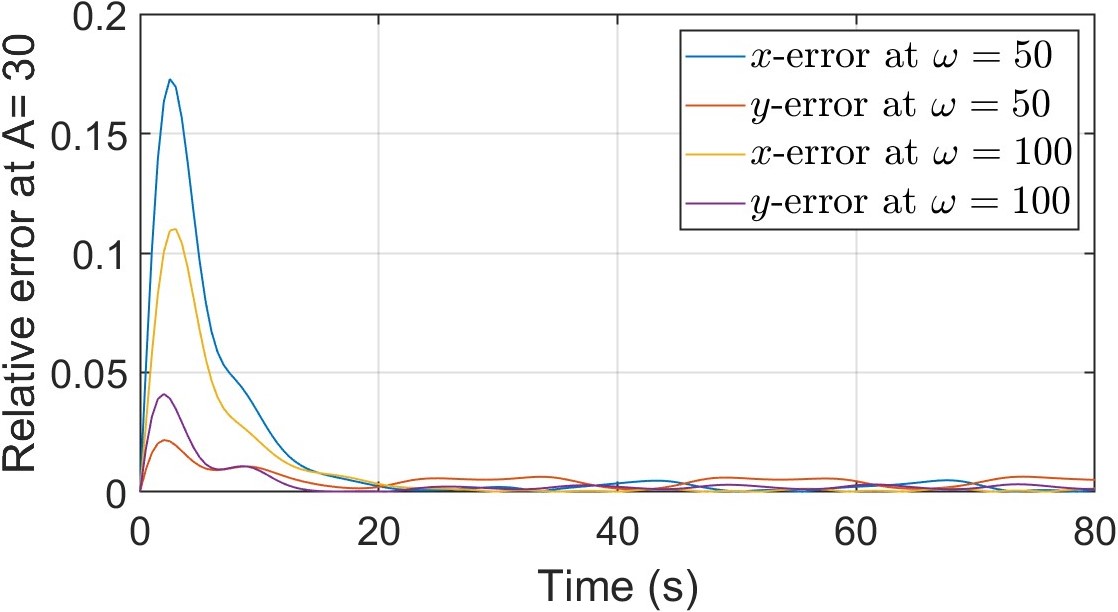} }}
   \\
       \subfloat[\centering Evolution of the error in the case of slow-varying disturbances is shown.]{{\includegraphics[width=0.8\columnwidth]{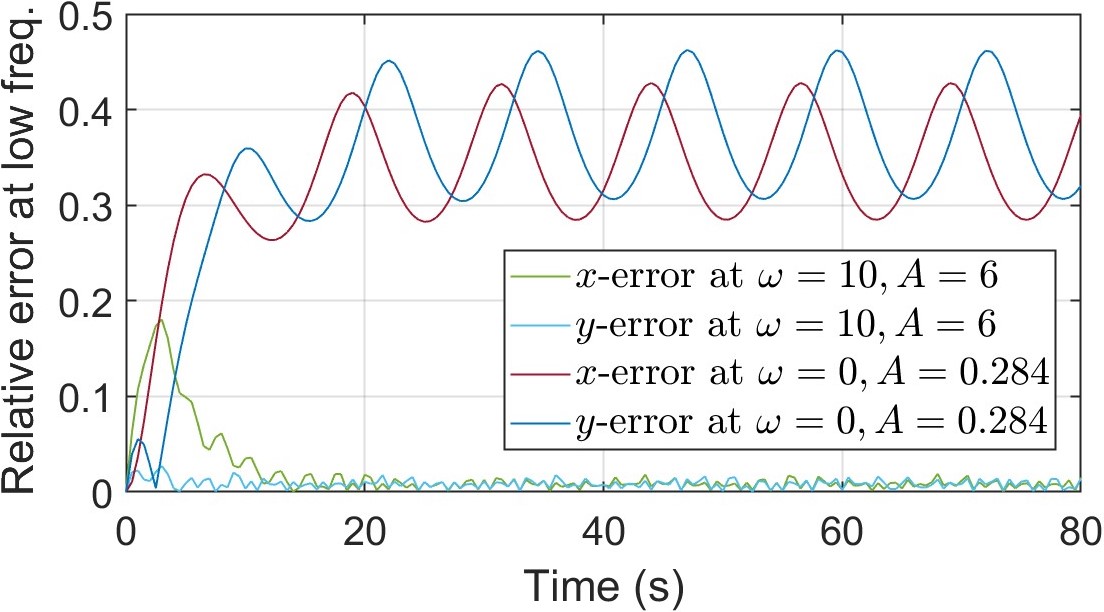} }}\\
    \subfloat[\centering Evolution of error in orientation tracking.]{\includegraphics[width=0.8\columnwidth]{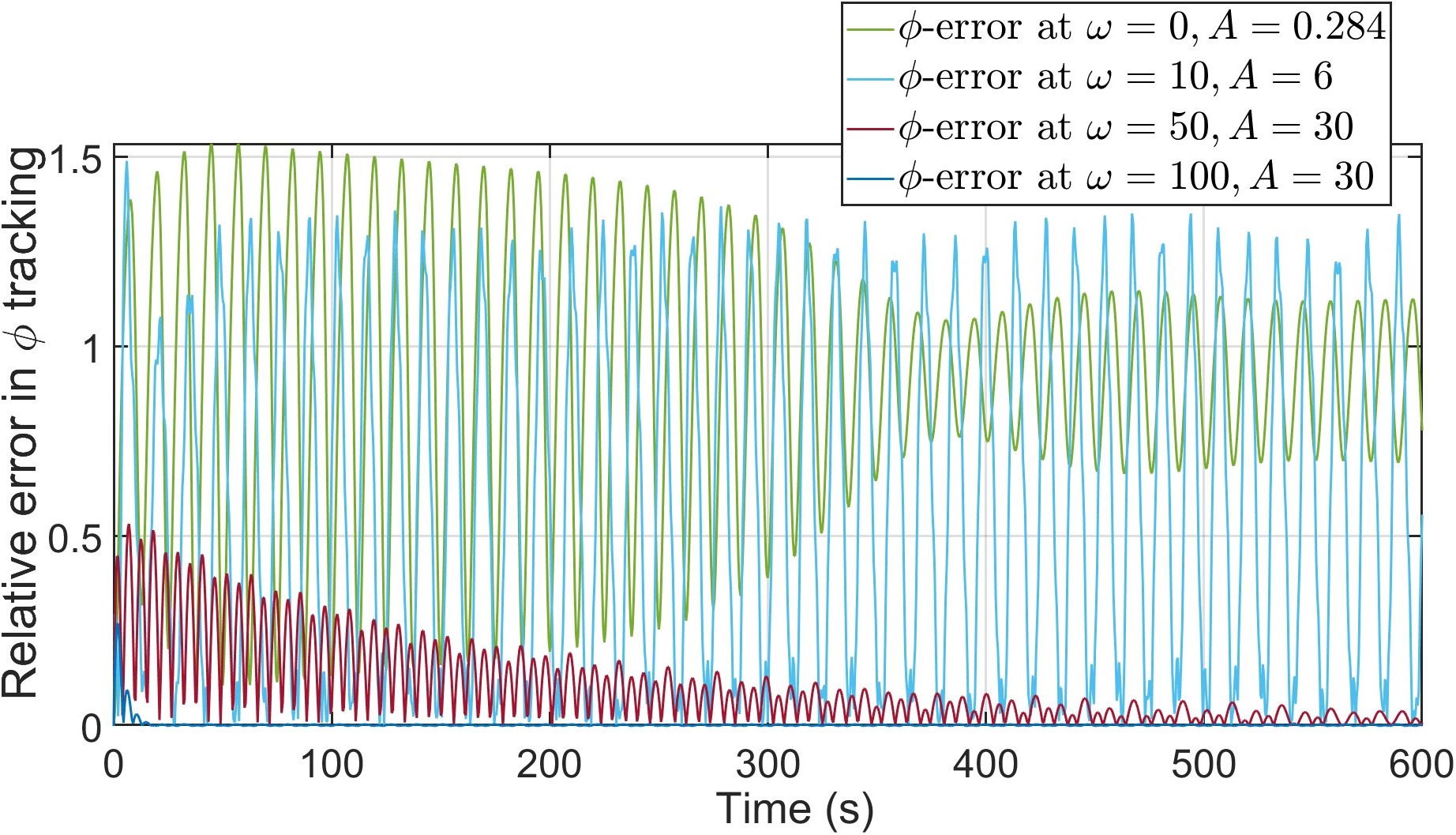}}
    \caption{Time evolution of output error is illustrated when disturbances having different frequencies enter the system output channels at the acceleration level.}
    \label{Dist_Periodic}
\end{figure}

In practice, external unmodeled forces may adversely alter the system behaviour and consequently damage the performance. In this section, We illustrate the effects of such disturbances $d(t)$ on the tracking performance. We consider periodic disturbances of the form
\begin{equation}
d(t)= A \sin(\omega \,t+\psi).
 \label{periodic}
\end{equation}
Constant disturbance of magnitude $A$ can be easily obtained by setting $(\omega=0, \psi=\pi/2)$ in~\eqref{periodic}. This disturbance is injected equally in each acceleration equation of the main platform pose, leading to the disturbance having a relative degree w.r.t that pose lower than the input relative degree (i.e. unmatched disturbance). In Fig.~\ref{Dist_Periodic}, the graph of the tracking error in each output component versus the allowed maximum disturbances across different frequencies the closed-loop system can withstand before running into singularity configurations is given.

\subsubsection{discussion}
As it can be noted from Table~\ref{Tab:pert_sep}, the system is sensitive to changes in the the masses more than in other parameters. Moreover, Fig.~\ref{Individ_pert} provides insights into how their perturbations affect the performance. A negative mismatch between the true value of platform's mass and the nominal value, used by the dynamic FBL controller, yields the highest impact on the performance compared to perturbations in the rest of the vehicle's parameters. Similar effect is observed when a positive variation in the propeller's link mass is considered. If both masses are perturbed to high values simultaneously, the error becomes larger. It is also worth noticing that perturbations in the inertial moment parameters seem to have little to no effect on the performance, which may be attributed to their small nominal values. In summary, it is much advisable that the measurement of masses composing the vehicle is as accurate as possible in order for the system to operate close to its nominal performance.

By examining how well the disturbance rejection capability of controller is, it is evident from Fig.~\ref{Dist_Periodic} that the system can successfully recover a good performance with smaller steady-state error, after transients vanish, when the disturbance is fast-varying compared to the case of constant or low frequency disturbance. This indicates that the capability of the closed-loop system to overcome the disturbance effect and recover the nominal performance weakens significantly with the decrease in frequency until a noticeable steady-state error appears. The error in orientation tracking exhibits this same phenomenon as position tracking error.
\section{Conclusion}
 We introduced a novel omnidirectional MAV concept that, contrarily to the existing solutions, has the minimum number of inputs equal to the DoFs of the main body and does not use reversible-thrust propellers. 
Moreover, thanks to the propellers aligning vertically at the steady state, no internal forces are produced at rest. We modeled the system using the EL approach, after which we discussed the omnidirectionality. We showed that the proposed vehicle is I/O feedback linearizable, w.r.t. an output from the main body pose, by a 4-dimensional dynamic extension, and the closed-loop equilibrium set of the extended system is stable. Finally, we simulated the closed-loop system response in a scenario where the vehicle performs a pose regulation motion. After this promising preliminary study in 2D, the research will be extended to the 3D setting. Designing robust controllers for this class of systems and carrying out laboratory experiments is left as future work.
\bibliographystyle{IEEEtran}
\bibliography{Bibliography}
\end{document}